\documentclass[11pt,a4paper]{amsart}

\usepackage{ifthen}
\newboolean{irmastyle}
\newboolean{draft}

\setboolean{irmastyle}{false}
\setboolean{draft}{false}

\ifthenelse{\boolean{draft}}{\usepackage{showkeys}}{}

\ifthenelse{\boolean{irmastyle}}{ 
\usepackage{amssymb,amsmath,latexsym,diagrams,graphicx}
\setcounter{page}{1}

\theoremstyle{definition}
 \newtheorem{dfn}{Definition}[section]
 \newtheorem{rem}[dfn]{Remark}
 \newtheorem{exa}[dfn]{Example}

\theoremstyle{plain}
 \newtheorem{pro}[dfn]{Proposition}
 \newtheorem{thm}[dfn]{Theorem}
 \newtheorem{cor}[dfn]{Corollary}
 \newtheorem{lem}[dfn]{Lemma}
 } 
{                                      
\usepackage{amsthm,amscd,times,amssymb,amsmath,diagrams,graphicx}

\newtheoremstyle{SmallCaps}{}{}{\itshape}{}{\textsc\bgroup}{.\egroup}{
 }{}
\newcounter{theorem}
\theoremstyle{SmallCaps}
\newtheorem{dfn}[theorem]{Definition}

\newtheorem{lem}[theorem]{Lemma}
\newtheorem{pro}[theorem]{Proposition}

\newtheorem{thm}[theorem]{Theorem}
} 

\newcommand{\cref}[1]{Corollary~\ref{#1}}

\newcommand{\tref}[1]{Theorem~\ref{#1}}
\def\please{\renewcommand\labelstyle{\scriptstyle}}
\def\R{\mathbb{R}}
\def\C{\mathbb{C}}
\def\N{\mathbb{N}}

\def\One{\mathbb{I}}
\def\F{\mathbb{F}}
\def\D{\mathcal{D}}

\def\ul{\underline}

\def\a{\alpha}

\ifthenelse{\boolean{draft}}{\date{Draft: \today}}{\date{}}

\def\be{\begin{equation*}}
\def\ee{\end{equation*}}
\def\bel{\begin{equation}}
\def\eel{\end{equation}}
\newcommand{\bea}{\begin{eqnarray}}
\newcommand{\eea}{\end{eqnarray}}
\newcommand{\beas}{\begin{eqnarray*}}
\newcommand{\eeas}{\end{eqnarray*}}

\begin{document}


\def\oct{                            
\parbox{10mm}{\begin{picture}(10,15) 
\put(4.5,12){$\bullet$}              
\put(1,5){$\bullet$} \put(8,5){$\circ$}
\put(5.5,13){\line(1,-2){3.3}} \put(5.5,13){\line(-1,-2){3.3}}
\put(1,2){$\ul t$}
\end{picture}}}

\def\oline{                          
\parbox{3mm}{\begin{picture}(5,10)   
\put(1,8){$\bullet$}                 
\put(1,1){$\bullet$} \put(2,9){\line(0,-1){7}}
\end{picture}}}

\def\olinephipsi{                    
\parbox{10mm}{\begin{picture}(10,10)  
\put(5,8){$$}                        
\put(5,1){$$} \put(7.25,9){\line(0,-1){7}}
\put(2,1){$\phi(\bullet)$} \put(1.5,8){$\psi(\bullet)$}
\end{picture}}}

\def\octline{                        
\parbox{3mm}{\begin{picture}(5,10)   
\put(1,8){$\ast   $}                 
\put(1,1){$\bullet$} \put(2,9){\line(0,-1){7}}
\end{picture}}}

\def\ocbline{                        
\parbox{3mm}{\begin{picture}(5,10)   
\put(1,8){$\bullet$}                 
\put(1,1){$\ast$} \put(2,9){\line(0,-1){7}}
\end{picture}}}

\def\octbline{                       
\parbox{3mm}{\begin{picture}(5,10)   
\put(1,8){$\ast   $}                 
\put(1,1){$\ast$} \put(2,9){\line(0,-1){7}}
\end{picture}}}

\def\oc{                     
\parbox{10mm}{\begin{picture}(10,10) 
\put(4.5,8){$\bullet$}               
\put(1,1){$\bullet$} \put(8,1){$\bullet$}
\put(5.5,9){\line(1,-2){3.5}} \put(5.5,9){\line(-1,-2){3.5}}
\end{picture}}}

\def\ocn{                    
\parbox{10mm}{\begin{picture}(10,10) 
\put(4.5,8){$\bullet$}               
\put(1,1){$\circ$} \put(8,1){$\circ$} \put(5.5,9){\line(1,-2){3.3}}
\put(5.5,9){\line(-1,-2){3.3}}
\end{picture}}}

\def\ocnr{                       
\parbox{10mm}{\begin{picture}(10,10) 
\put(4.5,8){$\bullet$}               
\put(1,1){$\bullet$} \put(8,1){$\circ$}
\put(5.5,9){\line(1,-2){3.3}} \put(5.5,9){\line(-1,-2){3.3}}
\end{picture}}}

\def\othree{                     
\parbox{10mm}{\begin{picture}(10,10) 
\put(4.5,8){$\bullet$}               
\put(1,1){$\bullet$} \put(4.5,1){$\bullet$} \put(8,1){$\bullet$}
\put(5.5,9){\line(1,-2){3.5}} \put(5.5,9){\line(-1,-2){3.5}}
\put(5.5,9){\line(0,-1){7}}
\end{picture}}}

\def\ofoura{                     
\parbox{13mm}{\begin{picture}(13,17) 
\put(8,15){$\bullet$}                
\put(4.5,8){$\bullet$}               
\put(8,8){$\bullet$}                 
\put(11.5,8){$\bullet$} \put(1,1){$\bullet$} \put(8,1){$\bullet$}
\put(9,16){\line(1,-2){3.5}} \put(9,16){\line(-1,-2){3.5}}
\put(5.5,9){\line(1,-2){3.5}} \put(5.5,9){\line(-1,-2){3.5}}
\put(9,16){\line(0,-1){7}}
\end{picture}}}

\def\occ{                     
\parbox{10mm}{\begin{picture}(10,17)  
\put(4.5,15){$\bullet$}               
\put(1,8){$\bullet$}                  
\put(8,8){$\bullet$}                  
\put(1,1){$\bullet$} \put(5.5,16){\line(1,-2){3.5}}
\put(5.5,16){\line(-1,-2){3.5}} \put(2,9){\line(0,-1){7}}
\end{picture}}}

\def\ofork{                           
\parbox{10mm}{\begin{picture}(10,17)  
\put(4.5,15){$\bullet$}               
\put(4.5,8){$\bullet$}                
\put(8,1){$\bullet$}                  
\put(1,1){$\bullet$} \put(5.5,9){\line(0,1){7}}
\put(5.5,9){\line(-1,-2){3.5}} \put(5.5,9){\line(1,-2){3.5}}
\end{picture}}}

\def\olongline{                       
\parbox{3mm}{\begin{picture}(3,17)    
\put(1,15){$\bullet$}                 
\put(1,8){$\bullet$}                  
\put(1,1){$\bullet$}                  
\put(2,2){\line(0,1){14}}
\end{picture}}}

\def\olonglongline{                   
\parbox{3mm}{\begin{picture}(3,24)    
\put(1,22){$\bullet$}                 
\put(1,15){$\bullet$}                 
\put(1,8){$\bullet$}                  
\put(1,1){$\bullet$}                  
\put(2,2){\line(0,1){21}}             
\end{picture}}}

\def\ocleft{                         
\parbox{10mm}{\begin{picture}(10,10) 
\put(4.5,8){$\bullet$}               
\put(1,1){$\ast$} \put(8,1){$\bullet$} \put(5.5,9){\line(1,-2){3.5}}
\put(5.5,9){\line(-1,-2){3.5}}
\end{picture}}}

\def\ocright{                        
\parbox{10mm}{\begin{picture}(10,10) 
\put(4.5,8){$\bullet$}               
\put(1,1){$\bullet$} \put(8,1){$\ast$} \put(5.5,9){\line(1,-2){3.5}}
\put(5.5,9){\line(-1,-2){3.5}}
\end{picture}}}

\def\ocleftright{                    
\parbox{10mm}{\begin{picture}(10,10) 
\put(4.5,8){$\bullet$}               
\put(1,1){$\ast$} \put(8,1){$\ast$} \put(5.5,9){\line(1,-2){3.5}}
\put(5.5,9){\line(-1,-2){3.5}}
\end{picture}}}

\def\octop{                          
\parbox{10mm}{\begin{picture}(10,10) 
\put(4.5,8){$\ast$}                  
\put(1,1){$\bullet$} \put(8,1){$\bullet$}
\put(5.5,9){\line(1,-2){3.5}} \put(5.5,9){\line(-1,-2){3.5}}
\end{picture}}}

\def\ocntop{                          
\parbox{10mm}{\begin{picture}(10,10) 
\put(4.5,8){$\ast$}                  
\put(1,1){$\circ$} \put(8,1){$\circ$} \put(5.5,9){\line(1,-2){3.3}}
\put(5.5,9){\line(-1,-2){3.3}}
\end{picture}}}

\def\oclefttop{                      
\parbox{10mm}{\begin{picture}(10,10) 
\put(4.5,8){$\ast$}                  
\put(1,1){$\ast$} \put(8,1){$\bullet$} \put(5.5,9){\line(1,-2){3.5}}
\put(5.5,9){\line(-1,-2){3.5}}
\end{picture}}}

\def\ocrighttop{                     
\parbox{10mm}{\begin{picture}(10,10) 
\put(4.5,8){$\ast$}                  
\put(1,1){$\bullet$} \put(8,1){$\ast$} \put(5.5,9){\line(1,-2){3.5}}
\put(5.5,9){\line(-1,-2){3.5}}
\end{picture}}}

\def\ocleftrighttop{                 
\parbox{10mm}{\begin{picture}(10,10) 
\put(4.5,8){$\ast$}                  
\put(1,1){$\ast$} \put(8,1){$\ast$} \put(5.5,9){\line(1,-2){3.5}}
\put(5.5,9){\line(-1,-2){3.5}}
\end{picture}}}

\def\olone{                          
\parbox{3mm}{\begin{picture}(5,10)   
\put(1,8){$\bullet_1$}               
\put(1,1){$\bullet$} \put(2,9){\line(0,-1){7}}
\end{picture}}}

\def\otwo{                         
\parbox{3mm}{\begin{picture}(5,10) 
\put(1,8){$\bullet_2$}             
\put(1,1){$\bullet$} \put(2,9){\line(0,-1){7}}
\end{picture}}}

\def\ocnp{                             
\parbox{16mm}{\begin{picture}(16,17)   
\put(4.5,8){$\bullet$}                 
\put(1,1){$\circ$}                     
\put(8,1){$\circ$}                     
\put(5.5,9){\line(1,-2){3.3}} \put(5.5,9){\line(-1,-2){3.3}}
\put(8,15){$\bullet$} \put(15,1){$\circ$}
\put(9,16){\line(-1,-2){3.5}} \put(9,16){\line(1,-2){6.7}}
\end{picture}}}

\def\ocnpo{                            
\parbox{16mm}{\begin{picture}(16,17)   
\put(4.5,8){$\bullet$}                 
\put(1,1){$\bullet$}                   
\put(8,1){$\bullet$}                   
\put(5.5,9){\line(1,-2){3.3}} \put(5.5,9){\line(-1,-2){3.3}}
\put(8,15){$\bullet$} \put(15,1){$\bullet$}
\put(9,16){\line(-1,-2){3.5}} \put(9,16){\line(1,-2){6.7}}
\end{picture}}}

\def\ocnplr{                           
\parbox{16mm}{\begin{picture}(16,17)   
\put(4.5,8){$\bullet_{12}$}            
\put(1,1){$\bullet_1$}                 
\put(8,1){$\bullet_2$}                 
\put(5.5,9){\line(1,-2){3.5}} \put(5.5,9){\line(-1,-2){3.5}}
\put(8,15){$\bullet_{123}$} \put(15,1){$\bullet_3$}
\put(9,16){\line(-1,-2){3.5}} \put(9,16){\line(1,-2){7}}
\end{picture}}}

\def\ocnprl{                           
\parbox{16mm}{\begin{picture}(16,17)   
\put(11.5,8){$\bullet_{23}$}           
\put(1,1){$\bullet_1$}                 
\put(8,1){$\bullet_2$}                 
\put(12.5,9){\line(1,-2){3.5}} \put(12.5,9){\line(-1,-2){3.5}}
\put(8,15){$\bullet_{123}$} \put(15,1){$\bullet_3$}
\put(9,16){\line(1,-2){3.5}} \put(9,16){\line(-1,-2){7}}
\end{picture}}}

\def\ocnpt{                            
\parbox{16mm}{\begin{picture}(16,17)   
\put(4.5,8){$\bullet$}                 
\put(1,1){$\circ$}                     
\put(8,1){$\circ$}                     
\put(5.5,9){\line(1,-2){3.3}} \put(5.5,9){\line(-1,-2){3.3}}
\put(8,15){$\ast$} \put(15,1){$\circ$} \put(9,16){\line(-1,-2){3.5}}
\put(9,16){\line(1,-2){6.7}}
\end{picture}}}

\def\ocnptb{                           
\parbox{16mm}{\begin{picture}(16,17)   
\put(4.5,8){$\ast   $}                 
\put(1,1){$\circ$}                     
\put(8,1){$\circ$}                     
\put(5.5,9){\line(1,-2){3.3}} \put(5.5,9){\line(-1,-2){3.3}}
\put(8,15){$\ast$} \put(15,1){$\circ$} \put(9,16){\line(-1,-2){3.5}}
\put(9,16){\line(1,-2){6.7}}
\end{picture}}}

\def\ocnpb{                            
\parbox{16mm}{\begin{picture}(16,17)   
\put(4.5,8){$\ast   $}                 
\put(1,1){$\circ$}                     
\put(8,1){$\circ$}                     
\put(5.5,9){\line(1,-2){3.3}} \put(5.5,9){\line(-1,-2){3.3}}
\put(8,15){$\bullet$} \put(15,1){$\circ$}
\put(9,16){\line(-1,-2){3.5}} \put(9,16){\line(1,-2){6.7}}
\end{picture}}}
\def\osix{                              
\parbox{37mm}{\begin{picture}(37,24)    
\put(4.5,8){$\bullet$}                  
\put(1,1){$\bullet$}                    
\put(8,1){$\bullet$}                    
\put(5.5,9){\line(1,-2){3.5}}           
\put(5.5,9){\line(-1,-2){3.5}} \put(18.5,8){$\bullet$}
\put(15,1){$\bullet$} \put(22,1){$\bullet$}
\put(19.5,9){\line(1,-2){3.5}} \put(19.5,9){\line(-1,-2){3.5}}
\put(32.5,8){$\bullet$} \put(29,1){$\bullet$} \put(36,1){$\bullet$}
\put(33.5,9){\line(1,-2){3.5}} \put(33.5,9){\line(-1,-2){3.5}}
\put(18.5,22){$\bullet$} \put(11.5,15){$\bullet$}
\put(19.5,9){\line(-1,1){7}} \put(12.5,16){\line(-1,-1){7}}
\put(19.5,23){\line(-1,-1){7}} \put(19.5,23){\line(1,-1){14}}
\end{picture}}}

\def\osixn{                 
\parbox{37mm}{\begin{picture}(37,24)    
\put(4.5,8){$\bullet$}                  
\put(1,1){$\circ$}                      
\put(8,1){$\circ$}                      
\put(5.5,9){\line(1,-2){3.3}}           
\put(5.5,9){\line(-1,-2){3.3}} \put(18.5,8){$\bullet$}
\put(15,1){$\circ$} \put(22,1){$\circ$}
\put(19.5,9){\line(1,-2){3.3}} \put(19.5,9){\line(-1,-2){3.3}}
\put(32.5,8){$\bullet$} \put(29,1){$\circ$} \put(36,1){$\circ$}
\put(33.5,9){\line(1,-2){3.3}} \put(33.5,9){\line(-1,-2){3.3}}
\put(18.5,22){$\bullet$} \put(19.5,23){\line(0,-1){14}}
\put(12.5,16){\line(-1,-1){7}} \put(19.5,23){\line(-1,-1){7}}
\put(19.5,23){\line(1,-1){14}}
\end{picture}}}

\def\osixend{              
\parbox{37mm}{\begin{picture}(37,24)    
\put(4.5,8){$\bullet$}                  
\put(1,1){$\bullet$}                    
\put(8,1){$\bullet$}                    
\put(5.5,9){\line(1,-2){3.5}}           
\put(5.5,9){\line(-1,-2){3.5}} \put(18.5,8){$\bullet$}
\put(15,1){$\bullet$} \put(22,1){$\bullet$}
\put(19.5,9){\line(1,-2){3.5}} \put(19.5,9){\line(-1,-2){3.5}}
\put(32.5,8){$\bullet$} \put(29,1){$\bullet$} \put(36,1){$\bullet$}
\put(33.5,9){\line(1,-2){3.5}} \put(33.5,9){\line(-1,-2){3.5}}
\put(18.5,22){$\bullet$} \put(19.5,23){\line(0,-1){14}}
\put(12.5,16){\line(-1,-1){7}} \put(19.5,23){\line(-1,-1){7}}
\put(19.5,23){\line(1,-1){14}}
\end{picture}}}

\def\osixnum{               
\parbox{80mm}{\begin{picture}(80,24)    
\put(4.5,8){$\bullet_{12}$}             
\put(1,1){$\bullet_1$}                  
\put(8,1){$\bullet_2$}                  
\put(5.5,9){\line(1,-2){3.5}}           
\put(5.5,9){\line(-1,-2){3.5}} \put(18.5,8){$\bullet_{34}$}
\put(15,1){$\bullet_3$} \put(22,1){$\bullet_4$}
\put(19.5,9){\line(1,-2){3.5}} \put(19.5,9){\line(-1,-2){3.5}}
\put(32.5,8){$\bullet_{56}$} \put(29,1){$\bullet_5$}
\put(36,1){$\bullet_6$} \put(33.5,9){\line(1,-2){3.5}}
\put(33.5,9){\line(-1,-2){3.5}} \put(11.5,15){$\bullet_{1234}$}
\put(18.5,22){$\bullet_{123456}$} \put(12.5,16){\line(1,-1){7}}
\put(12.5,16){\line(-1,-1){7}} \put(19.5,23){\line(-1,-1){7}}
\put(19.5,23){\line(1,-1){14}} \put(50,1){- level 0} \put(50,8){-
level 1} \put(50,15){- level 2} \put(50,22){- level 3}
\end{picture}}}

\def\ocfive{                         
\parbox{37mm}{\begin{picture}(37,24) 
\put(4.5,8){$\ast$}                  
\put(1,1){$\circ$}                   
\put(8,1){$\circ$}                   
\put(5.5,9){\line(1,-2){3.3}}        
\put(5.5,9){\line(-1,-2){3.3}}       
\put(18.5,8){$\ast$} \put(15,1){$\circ$} \put(22,1){$\circ$}
\put(19.5,9){\line(1,-2){3.3}} \put(19.5,9){\line(-1,-2){3.3}}
\put(32.5,8){$\bullet$} \put(29,1){$\circ$} \put(36,1){$\circ$}
\put(33.5,9){\line(1,-2){3.3}} \put(33.5,9){\line(-1,-2){3.3}}
\put(11.5,15){$\bullet$} \put(18.5,22){$\ast$}
\put(12.5,16){\line(1,-1){7}} \put(12.5,16){\line(-1,-1){7}}
\put(19.5,23){\line(-1,-1){7}} \put(19.5,23){\line(1,-1){14}}
\end{picture}}}

\def\ocful{
\parbox{16mm}{\begin{picture}(16,15)
\put(8,12){$\bullet$} \put(1,5){$\bullet$} \put(15,5){$\bullet$}
\put(9,13){\line(1,-1){7}} \put(9,13){\line(-1,-1){7}}
\put(9,13){\line(1,-2){3.5}} \put(9,13){\line(0,-1){7}}
\put(9,13){\line(-1,-2){3.5}} \put(1,2){${\ul t}_1$}
\put(7,2){$\ldots$} \put(15,2){${\ul t}_n$}
\end{picture}}}

\def\twoloop{
\parbox{20mm}{\begin{picture}(20,16)
\put(0,8){\line(1,0){4}} \put(4,8){\line(2,-1){16}}
\put(4,8){\line(2,1){16}} \put(12,4){\line(2,5){1.8}}
\put(16,14){\line(-2,-5){1.8}} \put(16,2){\line(-2,5){4}}
\put(12,4){\circle*{0.6}}
\end{picture}}}

\def\ocgammas{                                  
\parbox{10mm}{\begin{picture}(10,10)            
\put(3.5,8){$\bullet_{\gamma_1}$}               
\put(0,1){$\bullet_{\gamma_2}$}                 
\put(7,1){$\bullet_{\gamma_3}$} \put(4.5,9){\line(1,-2){3.5}}
\put(4.5,9){\line(-1,-2){3.5}}
\end{picture}}}

\def\olinegammaonetwo{                
\parbox{5mm}{\begin{picture}(5,10)    
\put(0,8){$\bullet_{\gamma_1}$}       
\put(0,1){$\bullet_{\gamma_2}$} \put(1,9){\line(0,-1){7}}
\end{picture}}}

\def\olinegammaonethree{              
\parbox{5mm}{\begin{picture}(5,10)    
\put(0,8){$\bullet_{\gamma_1}$}       
\put(0,1){$\bullet_{\gamma_3}$} \put(1,9){\line(0,-1){7}}
\end{picture}}}

\def\olinegammaonei{                  
\parbox{5mm}{\begin{picture}(5,10)    
\put(0,8){$\bullet_{\gamma_1}$}       
\put(0,1){$\bullet_{\gamma_i}$} \put(1,9){\line(0,-1){7}}
\end{picture}}}

\def\ocf{                                     
\parbox{16mm}{\begin{picture}(16,15)
\put(8,12){$\bullet$} 
\put(9,13){\line(1,-1){7}} \put(9,13){\line(-1,-1){7}}
\put(9,13){\line(1,-2){3.5}} \put(9,13){\line(0,-1){7}}
\put(9,13){\line(-1,-2){3.5}} \put(1,2){$\tau_1$}
\put(7,2){$\ldots$} \put(15,2){$\tau_n$}
\end{picture}}}

\def\oforknum{                        
\parbox{12mm}{\begin{picture}(12,17)  
\put(4.5,15){$\bullet_4$}             
\put(4.5,8){$\bullet_1$}              
\put(8,1){$\bullet_3$}                
\put(1,1){$\bullet_2$} \put(5.5,9){\line(0,1){7}}
\put(5.5,9){\line(-1,-2){3.5}} \put(5.5,9){\line(1,-2){3.5}}
\end{picture}}}

\def\oforknumhair{                    
\parbox{12mm}{\begin{picture}(12,24)  
\put(4.5,22){$\bullet^4$}             
\put(4.5,15){$\bullet_1$}             
\put(8,9){$\bullet_3$}                
\put(1,9){$\bullet_2$} \put(5.5,16){\line(0,1){7}}
\put(5.5,16){\line(-1,-2){3.5}} \put(5.5,16){\line(1,-2){3.5}}
\put(5.5,23){\line(-1,-2){3.5}} \put(5.5,23){\line(-2,-1){7}} 
\put(5.5,23){\line(1,-2){3.5}} \put(5.5,23){\line(2,-1){7}} 
\put(9,9){\line(-1,-2){3.5}} \put(9,9){\line(-1,-4){1.75}} 
\put(9,9){\line(1,-2){3.5}} \put(9,9){\line(1,-4){1.75}} 
\put(2,9){\line(-1,-2){3.5}} \put(2,9){\line(1,-4){1.75}} 
\put(2,9){\line(0,-1){7}}
\end{picture}}}


\def\phifull{\;\raisebox{-0.65cm}{\includegraphics[height=1.5cm]{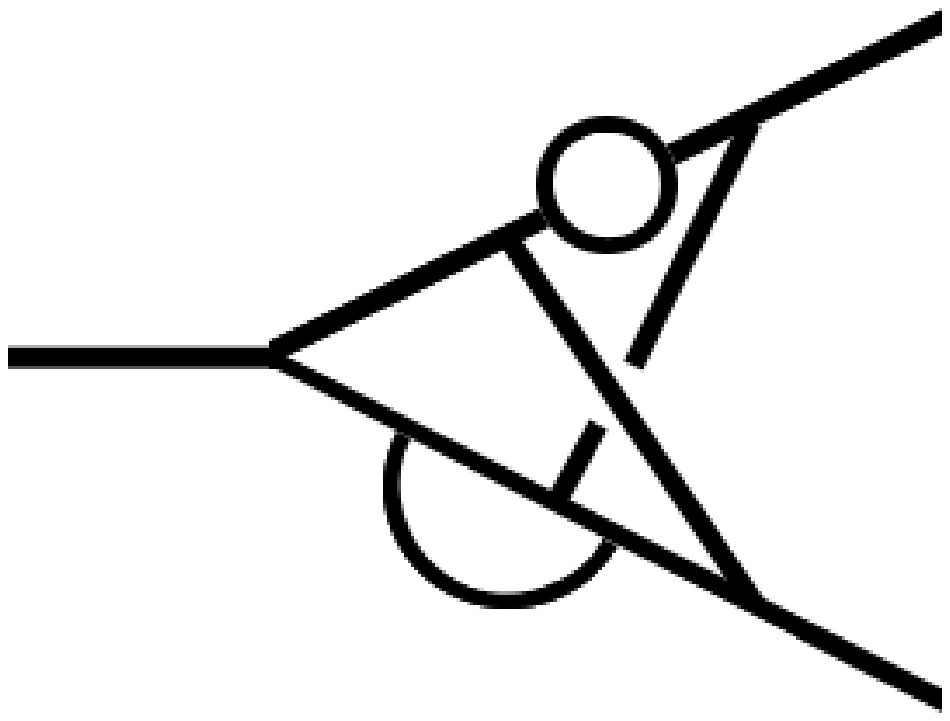}}\;}
\def\phitop{\;\raisebox{-0.65cm}{\includegraphics[height=1.5cm]{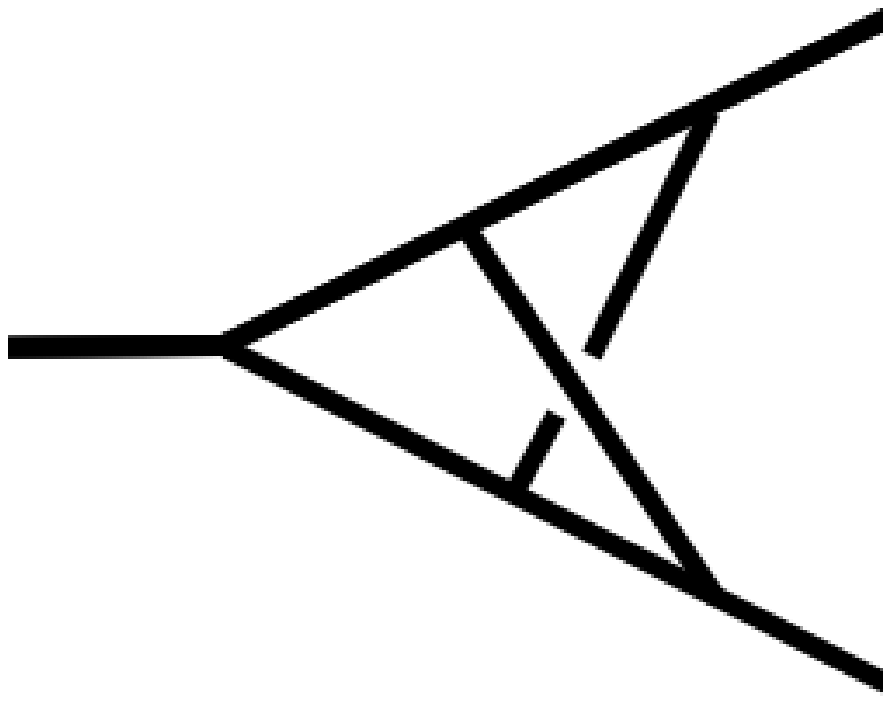}}\;}
\def\phileft{\;\raisebox{-0.25cm}{\includegraphics[height=0.75cm]{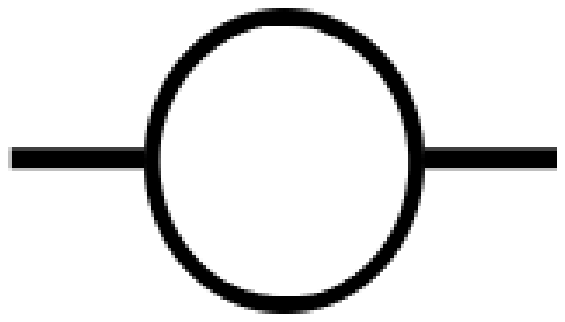}}\;}
\def\phiright{\;\raisebox{-0.40cm}{\includegraphics[height=1.00cm]{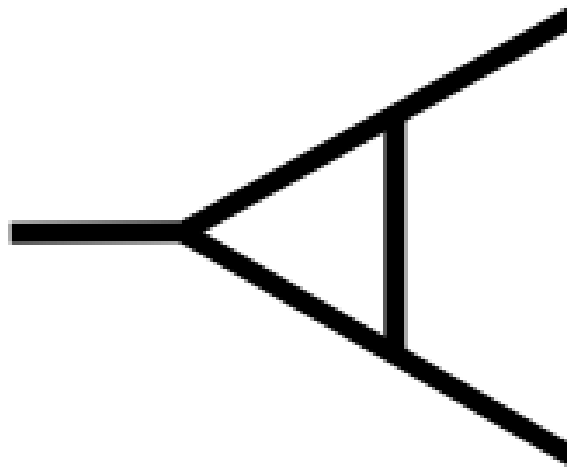}}\;}
\def\phidouble{\;\raisebox{-0.50cm}{\includegraphics[height=1.25cm]{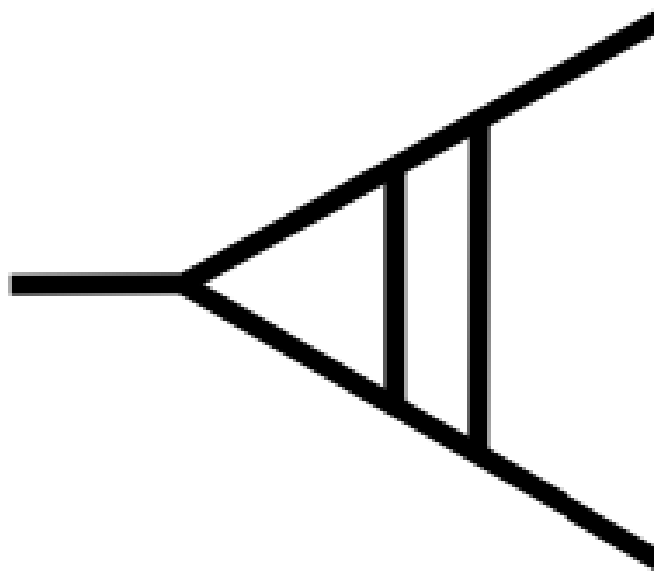}}\;}

\def\qedbfgga{\;\raisebox{-0.45cm}{\includegraphics[height=1cm]{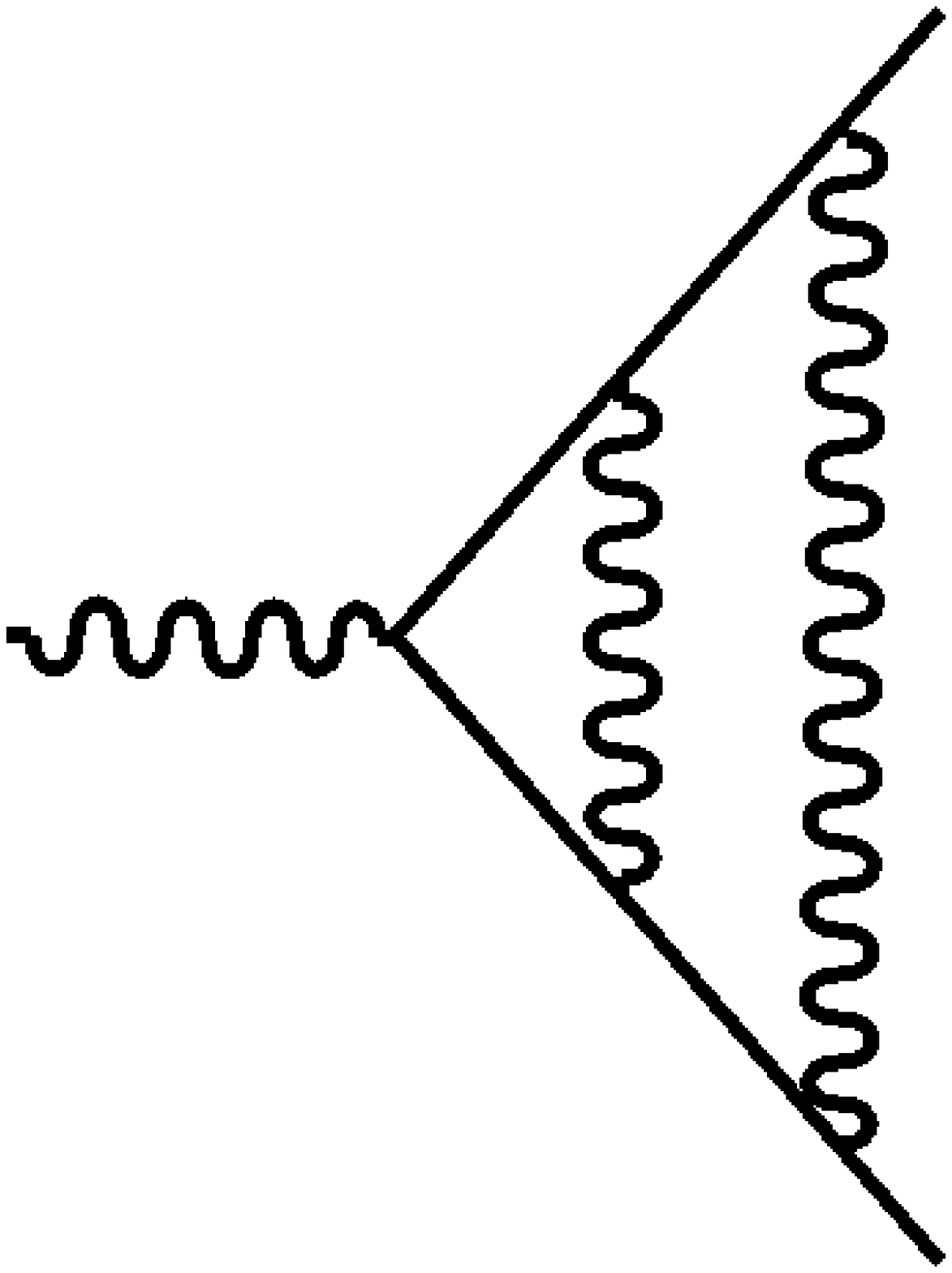}}\;}
\def\qedbfggb{\;\raisebox{-0.45cm}{\includegraphics[height=1cm]{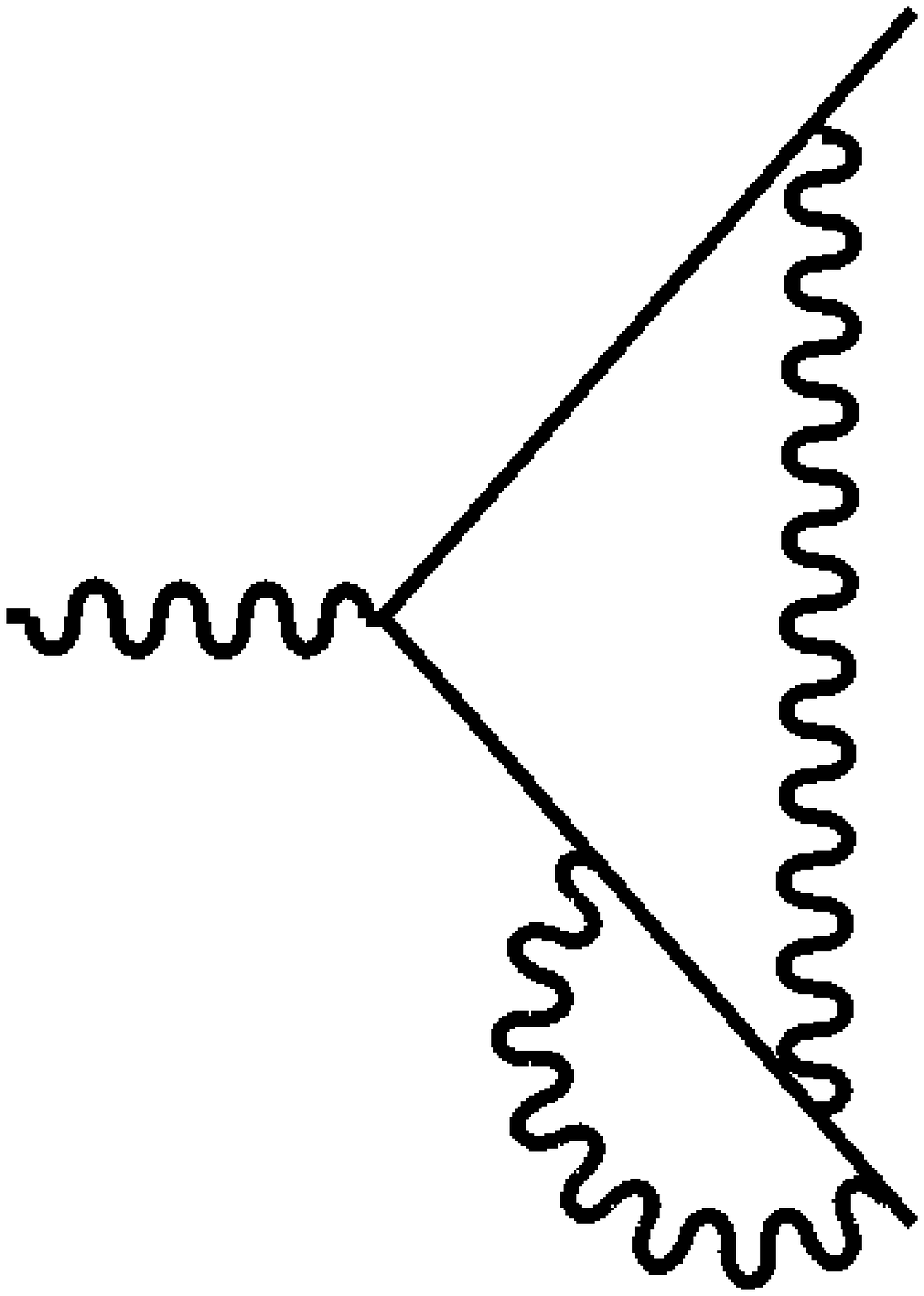}}\;}
\def\qedbfggc{\;\raisebox{-0.45cm}{\includegraphics[height=1cm]{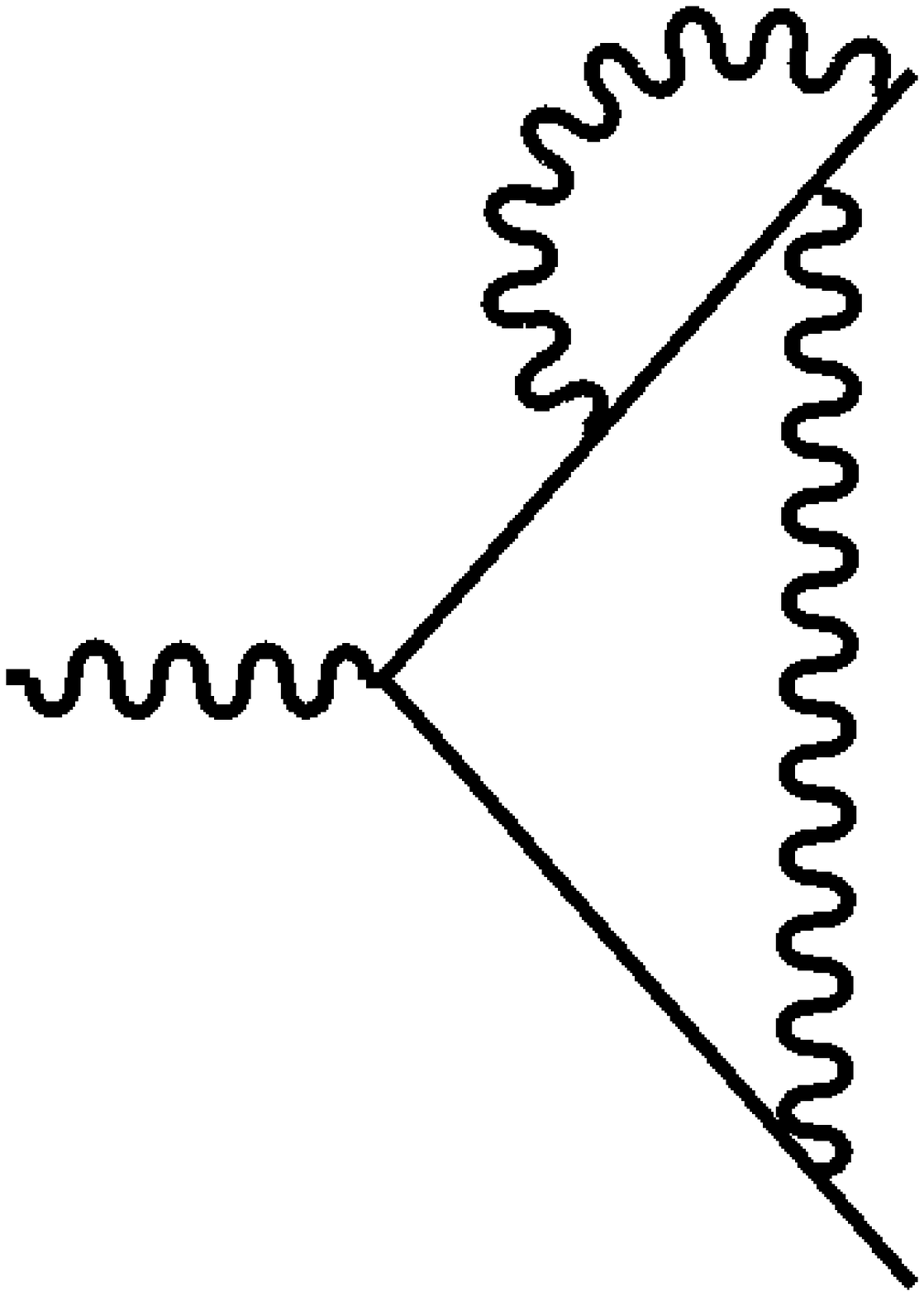}}\;}
\def\qedbfggd{\;\raisebox{-0.45cm}{\includegraphics[height=1cm]{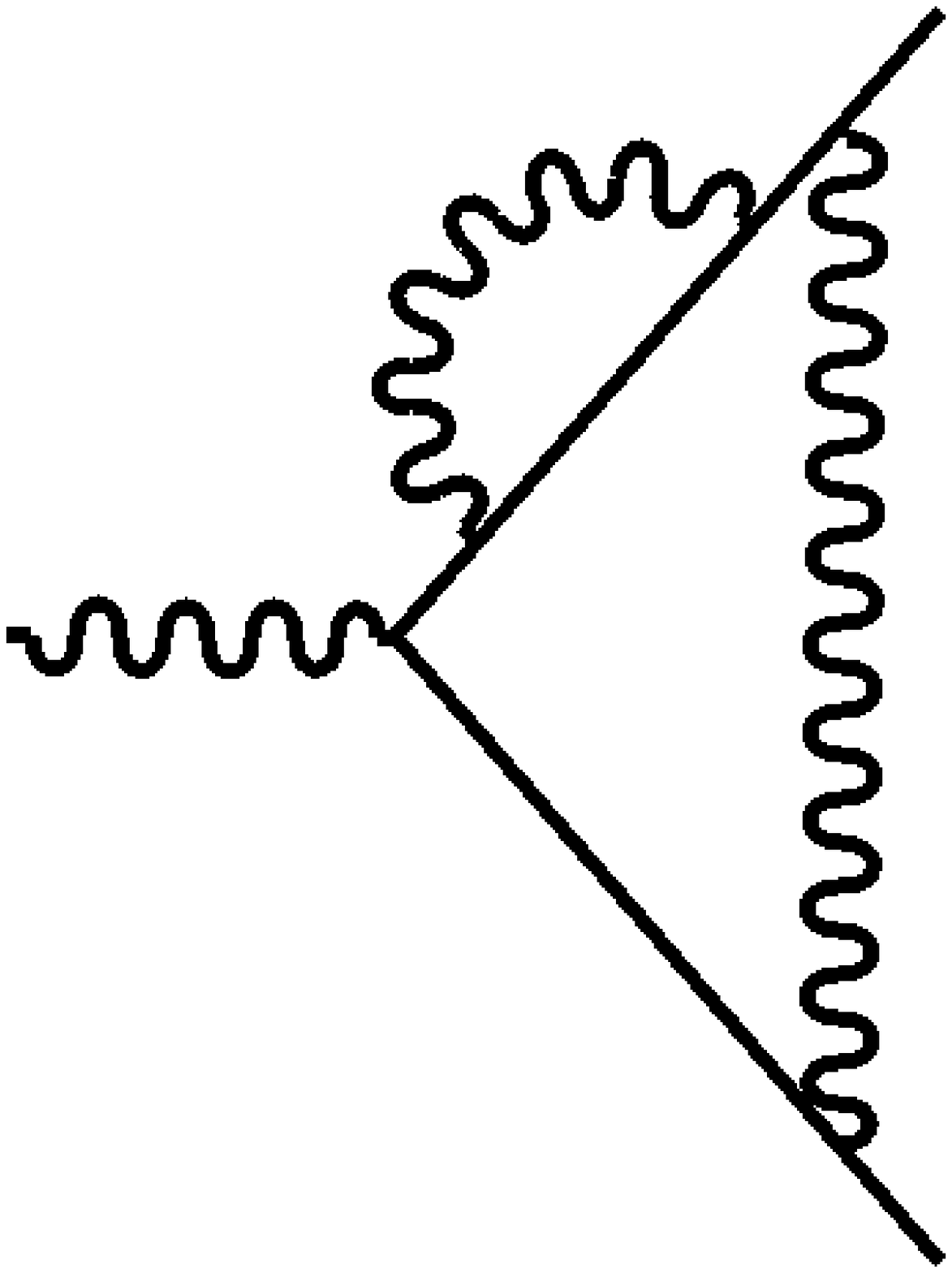}}\;}
\def\qedbfgge{\;\raisebox{-0.45cm}{\includegraphics[height=1cm]{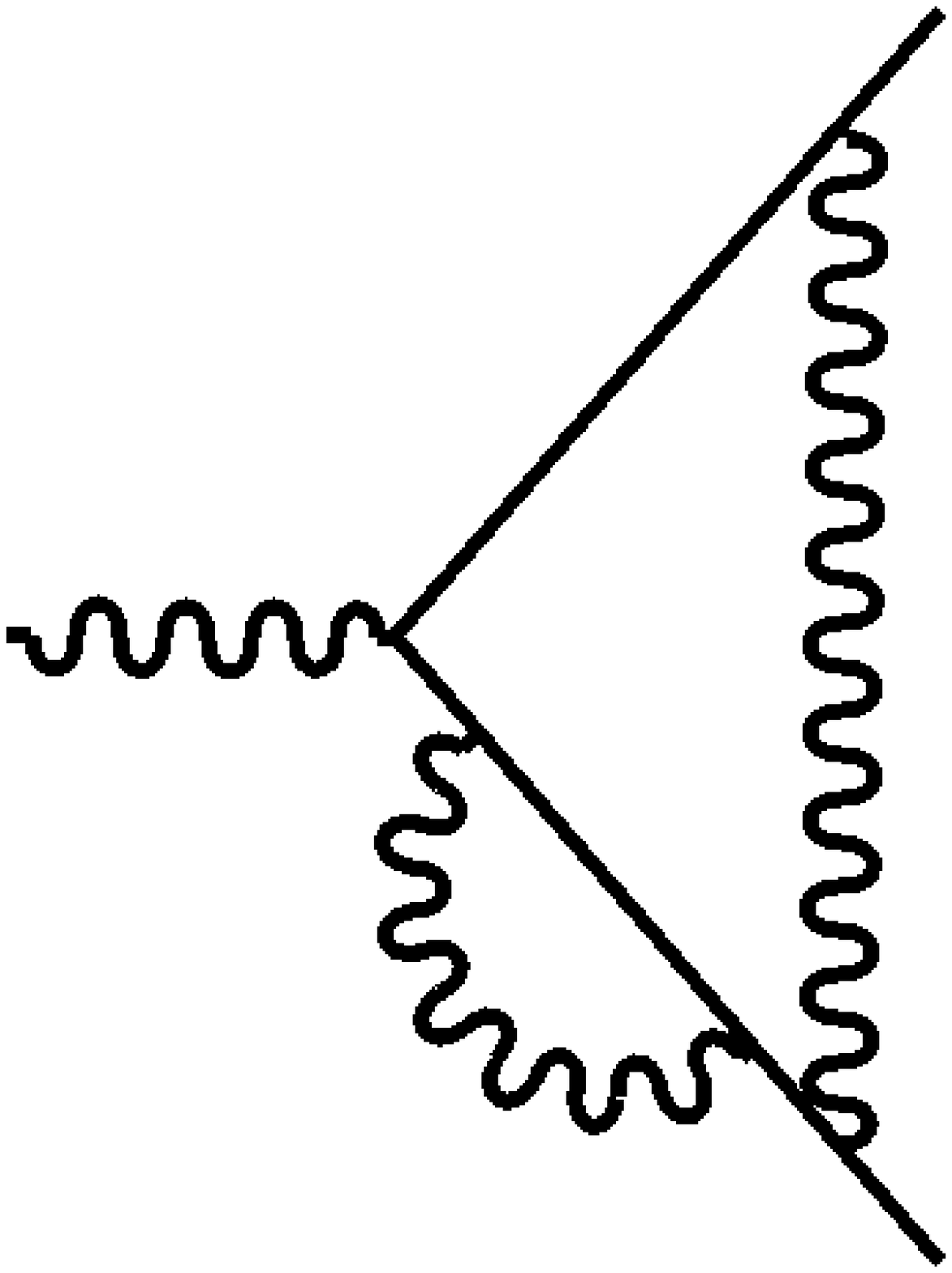}}\;}
\def\qedbfggf{\;\raisebox{-0.45cm}{\includegraphics[height=1cm]{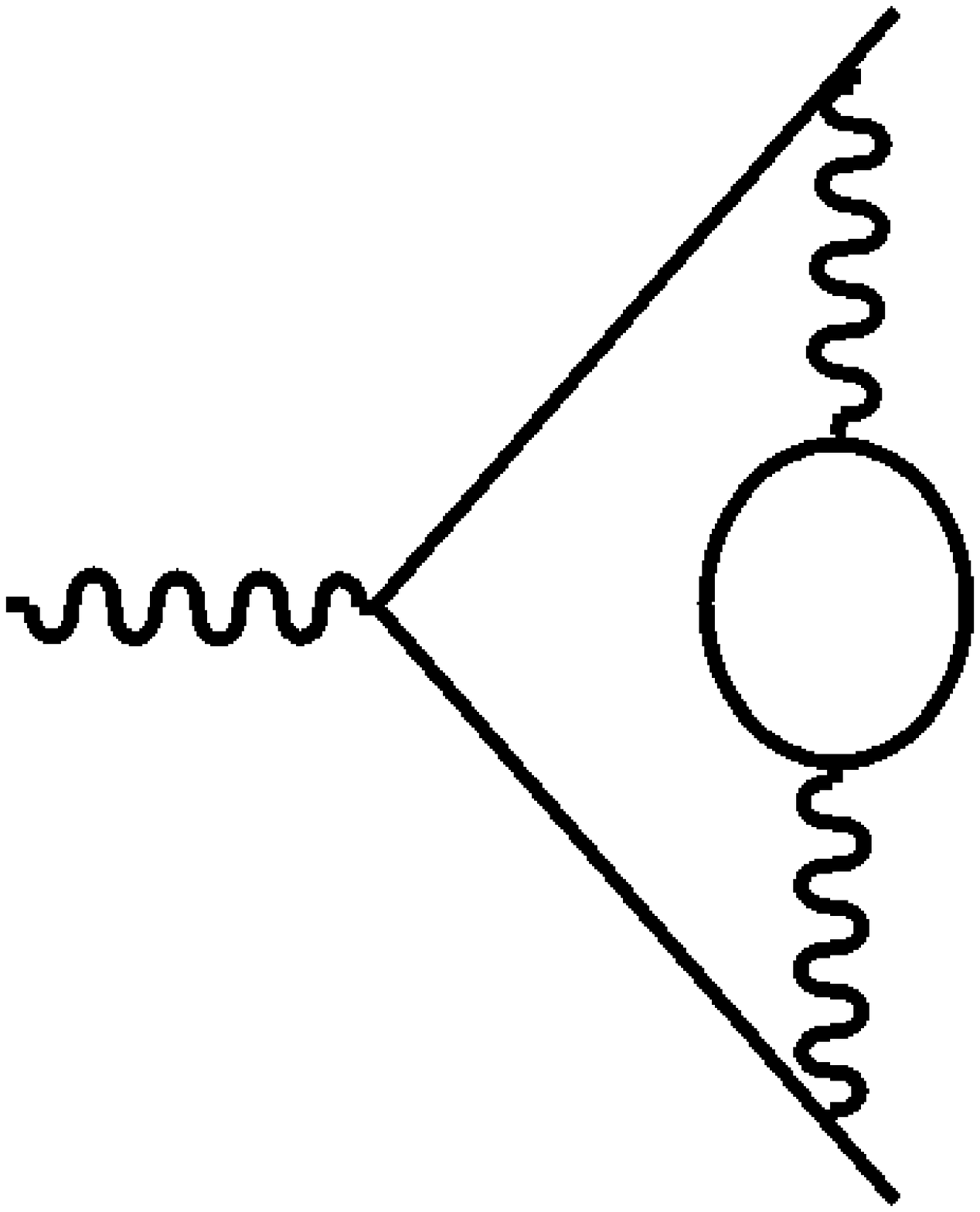}}\;}

\def\qedagffa{\;\raisebox{-0.15cm}{\includegraphics[height=6mm]{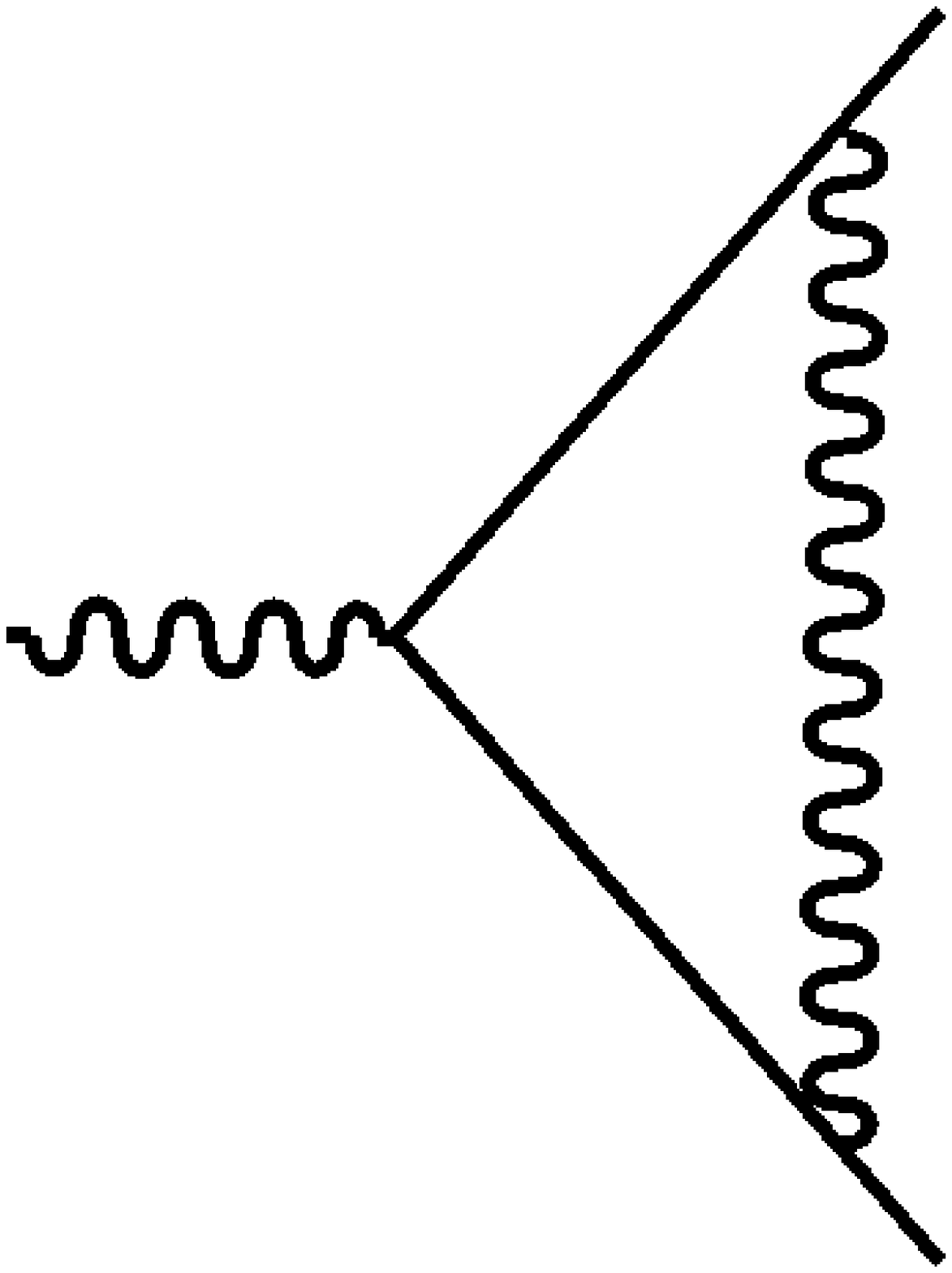}}\;}
\def\qedaffa{\;\raisebox{-0.10cm}{\includegraphics[height=4mm]{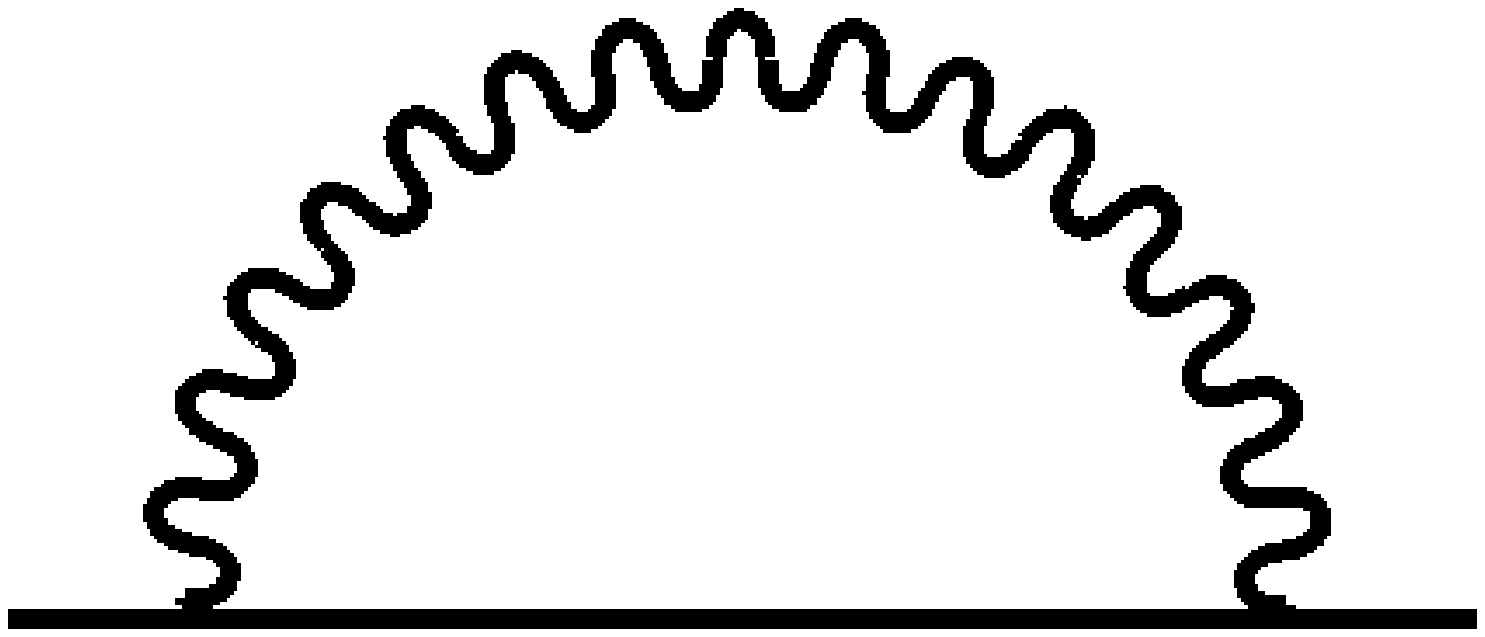}}\;}
\def\qedagga{\;\raisebox{-0.15cm}{\includegraphics[height=4mm]{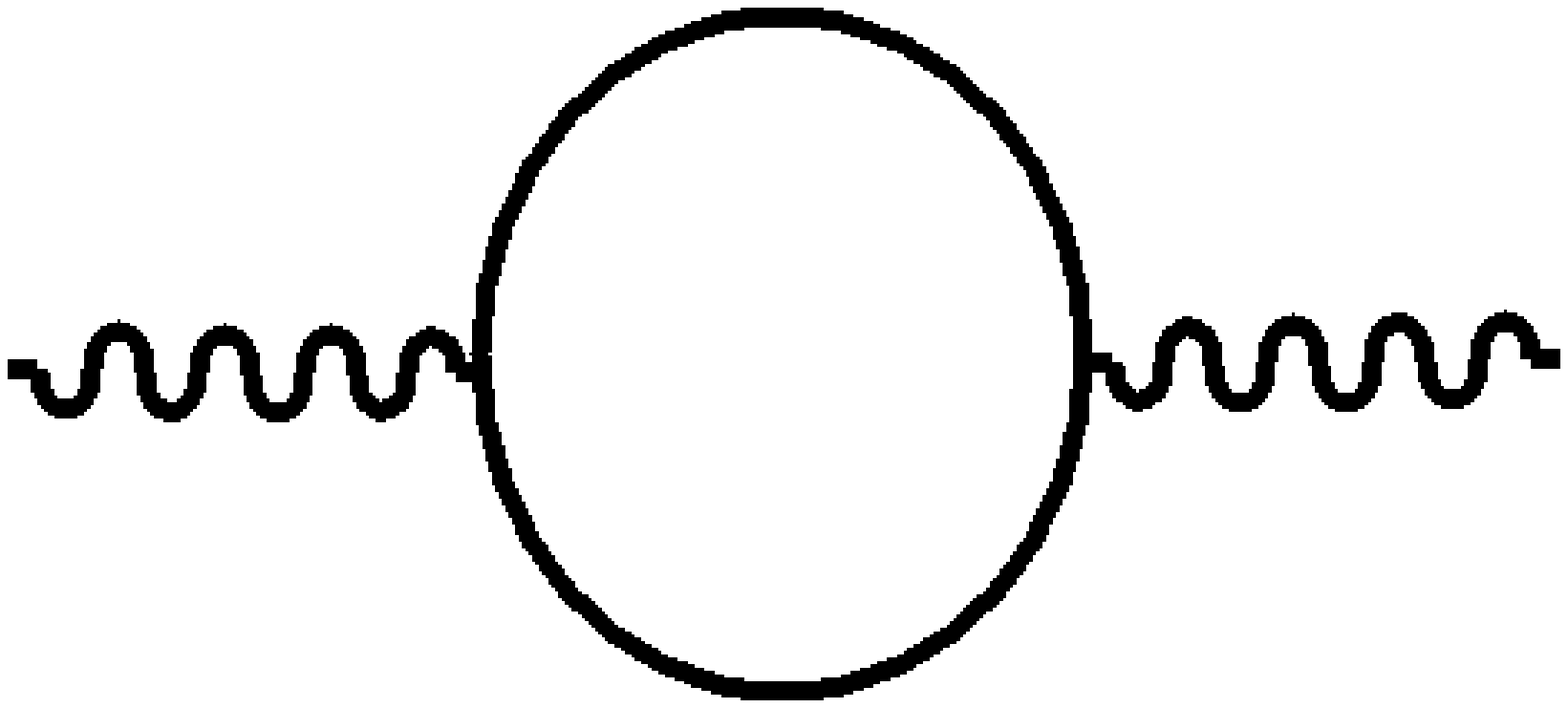}}\;}

\def\qedzff{\;\raisebox{+1mm}{\includegraphics[height=2mm]{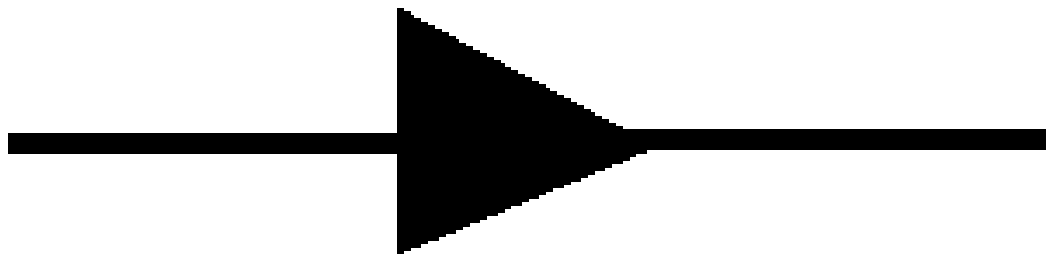}}\;}
\def\qedzggb{\;\raisebox{+1mm}{\includegraphics[height=2mm]{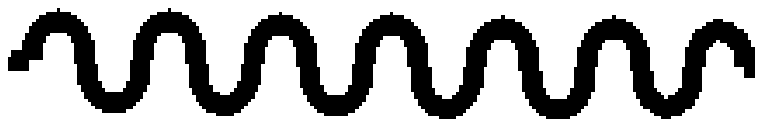}}\;}
\def\qedzgff{\;\raisebox{-2mm}{\includegraphics[height=6mm]{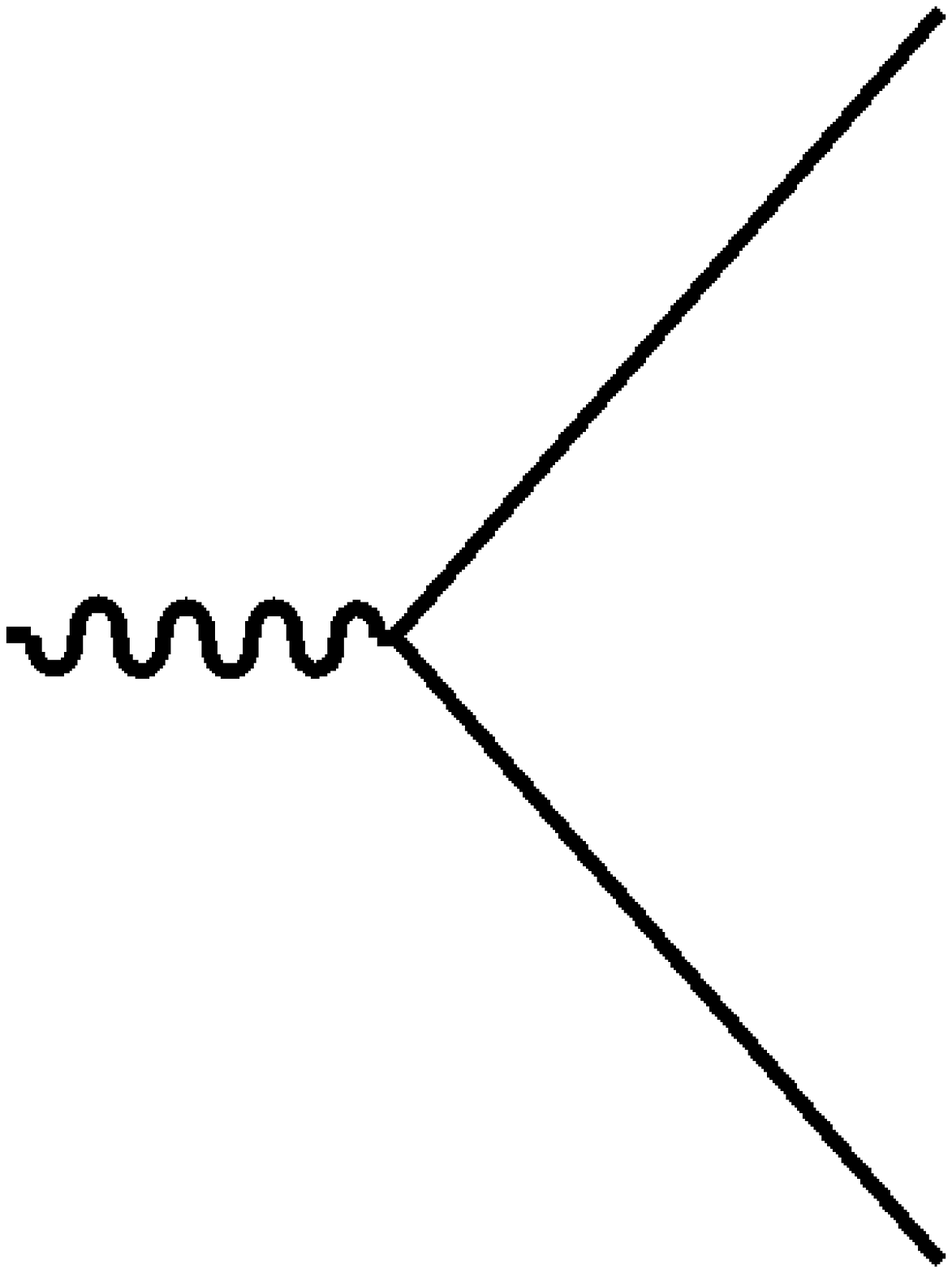}}\;}



\ifthenelse{\boolean{irmastyle}}{                                           
  \markboth{C.~Bergbauer and D.~Kreimer}{Hopf algebras in renormalization
  theory} 
\title{Hopf algebras in renormalization theory: Locality and Dyson-Schwinger equations from Hochschild cohomology}
\author{Christoph Bergbauer\thanks{FU Berlin and IHES. Supported by DFG grant VO 1272/1-1} and
Dirk Kreimer\thanks{CNRS at IHES and Boston University Center for
Mathematical Physics}}
\address{
Freie Universit\"at Berlin\\Institut f\"ur Mathematik
II\\Arnimallee 2-6\\D-14195 Berlin\\
email:\,\tt{bergbau@math.fu-berlin.de}
\\[4pt]
Institut des Hautes \'Etudes Scientifiques\\35 route de
Chartres\\F-91440 Bures-sur-Yvette\\ email:\,\tt{kreimer@ihes.fr}}

\maketitle

\begin{abstract}
In this review we discuss the relevance of the Hochschild
cohomology of renormalization Hopf algebras for local quantum
field theories and their equations of motion.
\end{abstract}



} { 
\title[Hopf algebras in renormalization theory]{Hopf algebras in renormalization theory: Locality and Dyson-Schwinger equations from Hochschild cohomology}
\author{C.~Bergbauer and D.~Kreimer}

\address{Freie Universit\"at Berlin\\Institut f\"ur Mathematik II\\Arnimallee 2-6\\D-14195 Berlin and \newline Institut des Hautes \'Etudes Scientifiques\\35 route de Chartres\\F-91440 Bures-sur-Yvette}
\email{bergbau@math.fu-berlin.de}

\address{CNRS at Institut des Hautes \'Etudes Scientifiques\\35 route de Chartres\\F-91440 Bures-sur-Yvette and
Center for Mathematical Physics, Boston University}
\email{kreimer@ihes.fr}

\begin{abstract}
In this review we discuss the relevance of the Hochschild
cohomology of renormalization Hopf algebras for local quantum
field theories and their equations of motion.
\end{abstract}

\maketitle

} 

\setcounter{tocdepth}{2} \tableofcontents


\section*{Introduction and acknowledgments}
The relevance of infinite dimensional Hopf and Lie algebras for the
understanding of local quantum field theory has been established in
the last couple of years. Here, we focus on the role of the
1-cocycles in the Hochschild cohomology of such renormalization Hopf
algebras.\\\\
After an introductory overview which recapitulates the well-known
Hopf algebra of rooted trees we exhibit the crucial connection
between 1-cocycles in the Hochschild cohomology of the Hopf
algebra, locality and the structure of the quantum equations of
motion. For the latter, we introduce combinatorial Dyson-Schwinger
equations and show that the perturbation series provides Hopf
subalgebras indexed only by the order of the perturbation. We then
discuss assorted applications of such equations which focus on the
notion of self-similarity and transcendence.\\\\
This paper is based on an overview talk given by one of us (D.\ K.),
extended by a more detailed exhibition of some useful mathematical
aspects of the Hochschild cohomology of the relevant Hopf algebras.
It is a pleasure to thank the organizers of the  {\em 75\`eme
Rencontre entre Physiciens Th\'eoriciens et Math\'ematiciens} for
organizing that enjoyable workshop. D.~K.\ thanks Karen Yeats for
discussions on the transcendental nature of DSEs. C.~B.\
acknowledges support by the Deutsche Forschungsgemeinschaft under grant VO 1272/1-1. He also
thanks Boston University and the IHES for hospitality.
\section{Rooted trees, Feynman graphs, Hochschild cohomology and local counterterms}
\subsection{Motivation}\label{ss:motivation}
Rooted trees store information about nested and disjoint
subdivergences of Feynman graphs in a natural way. This has been
used at least implicitly since Hepp's proof of the BPH subtraction
formula \cite{HeppCMP} and Zimmermann's forest formula
\cite{Zimmermann}. However it was only decades later that the
algebraic structure of the Bogoliubov recursion was elucidated by
showing that it is essentially given by the coproduct and the
corresponding antipode of a Hopf algebra on rooted trees
\cite{Kreimer,CK}. The same result can be formulated more directly
in terms of a very similar Hopf algebra on 1PI Feynman graphs
\cite{CK2}. We start with the description in terms of rooted trees
which serves as a universal
role model for all Hopf algebras of this kind.\\\\
For instance, the subdivergences of the $\phi^3$ diagram in six
spacetime dimensions
\begin{center}
$\Gamma=$\phifull
\end{center}
can be represented by the decorated tree
\begin{equation*}
\ocgammas
\end{equation*}
where
\begin{center}
$\gamma_1=$\phitop,$\quad\gamma_2=$\phileft,$\quad\gamma_3=$\phiright.
\end{center}
Additional labelling (which we do not care about here) would be
needed to keep track of the actual insertion places. However, since
one is ultimately interested in the sum of all Feynman graphs of a
given order in perturbation theory, for the purpose of the
Bogoliubov recursion all possible insertions of $\gamma_2$ and
$\gamma_3$ into $\gamma_1$ can be considered at the same time, when
due care is given to the resolution of graphs with overlapping
divergences into appropriate linear combinations of trees. \\\\
In a moment we will need the trees
\begin{equation*}
\olinegammaonetwo\;\mbox{ and }\;\olinegammaonethree
\end{equation*}
whose meaning should be clear: They represent the graph $\gamma_1$
for which
$\gamma_2$ or $\gamma_3,$ respectively, is suitably inserted. \\\\
Now how is $\Gamma$ renormalized? According to the Bogoliubov
recursion, the renormalized value is given by
\begin{eqnarray}\label{eq:Bogo}
\phi_R\left(\ocgammas\right) & := &
(id-R)\left(\phi\left(\ocgammas\right)-R\phi(\bullet_{\gamma_2})\phi\left(\olinegammaonethree\right)-\right.\nonumber\\
& & -R\phi(\bullet_{\gamma_3})\phi\left(\olinegammaonetwo\right)-
R\left(\phi(\bullet_{\gamma_2}\bullet_{\gamma_3})-\right.\\
&& -\biggl.\left.\phi(\bullet_{\gamma_2})R\phi(\bullet_{\gamma_3})-
\phi(\bullet_{\gamma_3})R\phi(\bullet_{\gamma_2})\right)\phi(\bullet_{\gamma_1})\biggr)\nonumber
\end{eqnarray}
where $\phi$ denotes the unrenormalized but possibly regularized (if
we do not renormalize on the level of the integrand) contribution of
the graph which a given tree represents. For example, in dimensional
regularization, $\phi$ is a map into the algebra
$V:=\C[\epsilon^{-1},\epsilon]]$ of Laurent series with finite pole
part. The map $R:V\rightarrow V$ is a renormalization scheme. For
example, the minimal subtraction scheme is obtained by defining $R$
to be the projector onto the proper pole part,
$R(\epsilon^k)=\epsilon^k$ if $k<0$ and $R(\epsilon^k)=0$ otherwise.
We emphasize though that the use of a regulator can be avoided by
defining a suitable renormalization scheme on the level of
integrands. Such an approach can then be directly formulated on the
level of Dyson--Schwinger equations, where the choice of a
renormalization scheme can be non-perturbatively given as the choice
of a boundary condition
for the accompanying integral equation.\\\\
Now consider the polynomial algebra $\mathcal{H}$ generated by all
decorated rooted trees of this kind. There is a coproduct on it
which disentangles trees into subtrees and thus divergences into
subdivergences, as will be discussed in subsection \ref{ss:trees}.
Using this coproduct $\Delta,$ the above algebra $\mathcal{H}$
becomes a Hopf algebra. Let $S$ be its antipode. By definition,
$S$ satisfies the recursive relation (in Sweedler's notation
$\Delta(x)=\One\otimes x+x\otimes\One+\tilde\sum x'\otimes x''$)
\begin{equation*}
S(x) = -x - \tilde\sum S(x')x''
\end{equation*}
It turns out that, if one similarly defines a ''twisted antipode``
$S_R^\phi$ as a map $\mathcal{H}\rightarrow V$ by $S_R^\phi(1)=1$
and
\begin{equation*}
S_R^\phi(x) = - R\left(\phi(x)+\tilde\sum
S^\phi_R(x')\phi(x'')\right),
\end{equation*}
then $S_R^\phi$ provides the counterterm and the convolution
$S_R^\phi\star\phi = m_V(S_R\otimes \phi)\Delta$ solves the
Bogoliubov recursion: $\phi_R = S_R^\phi\star\phi$
\cite{Kreimer,CK}. Using this algebraic approach to the
combinatorial intricacies of renormalization, many important
questions in perturbative and non-perturbative quantum field theory
can be treated from a convenient conceptual point of view, some of which will be reviewed
 in the following sections. \\\\
This picture translates rather easily to renormalization in
coordinate space \cite{BK}, as will be briefly discussed in subsection \ref{ss:locality}.\\\\
Before continuing the discussion of renormalization, we introduce
some key algebraic notions. We will come back to the example of the
graph $\Gamma$ later on.
\subsection{Basic definitions and notation}
Let $k$ be a field of characteristic zero. We consider
$k$-bialgebras $(A,m,\One,\Delta,\epsilon)$ that are graded
connected, that is
\begin{equation*}
A=\bigoplus_{n=0}^\infty A_n,\quad A_0\cong k,\quad A_mA_n\subseteq A_{m+n},\quad \Delta(A_n)\subseteq \bigoplus_{l+m=n} A_l\otimes A_m.
\end{equation*}
By abuse of notation, we write $\One$ both for the unit and the unit
map. Also, we sometimes consider $\epsilon$ as a map $A\rightarrow
A_0.$ We assume that $\Delta(\One)=\One\otimes\One.$ It follows that
$\epsilon(\One)=1$ while $\epsilon(A_n)=0$ for $n\neq 0.$ The kernel
of $\epsilon$ is called the augmentation ideal, and the map $P:
A\rightarrow A,$ $P=id-\epsilon,$ is called the projection onto the
augmentation ideal. The coproduct $\Delta$ gives rise to another
coassociative map: $\tilde\Delta,$ defined by
\begin{equation*}
\tilde\Delta(x) = \Delta(x)-\One\otimes x-x\otimes\One.
\end{equation*}
Recall that elements in the kernel of $\tilde\Delta$ are called
\emph{primitive.} We will occasionally use Sweedler's notation
$\Delta(x)=\sum x'\otimes x''$ and also
$\tilde\Delta(x)=\tilde\sum x'\otimes x''.$\\\\
It is a well known fact that connected graded bialgebras are Hopf
algebras. Indeed, the sequence defined by the recursive relation
\begin{equation}\label{eq:antipode}
S(x) = -x - \tilde\sum S(x')x''\mbox{ for }x\not\in A_0,\quad
S(\One)=\One
\end{equation}
converges in End$_k(A).$\\\\
For a coalgebra $(A,\Delta)$ and an algebra $(B,m)$, the vector
space $\operatorname{Hom}_k(A,B)$ of linear maps $A\rightarrow B$ is
equipped with a convolution product $\star$ by $(f,g)\mapsto f\star
g = m(f\otimes g)\Delta.$ Thus $(f\star g)(x)=\sum f(x')g(x'').$
Using the modified product $\star_P: (f,g)\mapsto f\star_P g =
m(f\otimes g)(id\otimes P)\Delta,$ equations (\ref{eq:antipode}) can
be rewritten
\begin{equation*}
S(x) = -(S\star_P id)(x)\mbox{ for }x\not\in A_0, \quad S(\One)=\One
\end{equation*}
which will be convenient later on.
\subsection{The Hopf algebra of rooted trees}\label{ss:trees}
Now we give a more detailed construction of the Hopf algebra
$\mathcal{H}$ of rooted trees \cite{Kreimer,CK} that is in the
center of all our considerations. An (undecorated, non-planar)
rooted tree is a connected contractible finite graph with a
distinguished vertex called the root. By convention, we will draw
the root on top. We are only interested in isomorphism classes of
rooted trees (an isomorphism of rooted trees being an isomorphism
of graphs which maps the root to the root) which we, by abuse of
language, simply call rooted trees again. As a graded algebra,
$\mathcal{H}$ is the free commutative algebra generated by trees
(including the empty tree which we consider the unit $\One$) with
the weight grading: the weight of a tree is the number of its
vertices. For instance, the trees of weight one to four are
\begin{equation*}
\bullet,\,\oline,\,\oc,\,\olongline,\,\othree,\,\occ,\,\ofork,\,\olonglongline.
\end{equation*}
A product of rooted trees is called a forest -- obviously the
weight of a forest is the sum of the weights of its trees. On
$\mathcal{H}$ a coproduct $\Delta$ is introduced by
\begin{equation}\label{cuts}
\Delta (\tau)=\One\otimes \tau+\tau\otimes\One+ \sum_{adm. c}
P_c(\tau)\otimes R_c(\tau)
\end{equation}
where the sum goes over all \emph{admissible cuts} of the tree
$\tau.$ By a cut of $\tau$ we mean a nonempty subset of the edges of
$\tau$ that are to be removed. The product of subtrees which ``fall
down'' upon removal of those edges is called the \emph{pruned part}
and is denoted $P_c(\tau),$ the part which remains connected with
the root is denoted $R_c(\tau).$ This makes sense only for certain
''admissible'' cuts: by definition, a cut $c(\tau)$ is admissible,
if for each leaf $l$ of $\tau$ it contains at most one edge on the
unique path from $l$ to the root. For instance,
\begin{eqnarray*}
\Delta\left(\ofork\right)&=&\ofork\otimes\One+\One\otimes\ofork+2\bullet\otimes\olongline+\\
&+&\bullet\bullet\otimes\oline+\oc\otimes\bullet.
\end{eqnarray*}
The coassociativity of $\Delta$ is shown in \cite{Kreimer}.
$\mathcal{H}$ is obviously not cocommutative. Since the coproduct is
compatible with the grading, $\mathcal{H}$ is a Hopf algebra. There
is an important linear endomorphism of $\mathcal{H},$ the grafting
operator $B_+$ defined as follows:
\begin{eqnarray}\label{eq:defbplus}
B_+(\One)&=&\bullet\nonumber\\
B_+(\tau_1 \ldots\tau_n)&=&\ocf\quad\mbox{for trees }\tau_i
\end{eqnarray}
In words: $B_+$ creates a new root and connects it with each root
of its argument. The special importance of $B_+$ will become
evident in subsection \ref{ss:Hochschild}:
$B_+:\mathcal{H}\rightarrow\mathcal{H}$ is a closed but not exact
Hochschild 1-cochain. \\\\
The Hopf algebra $\mathcal{H}$ is the dual of a Hopf algebra
considered earlier by Grossman and Larson \cite{GL}, see
\cite{FoissyBSM1}. It can also be described from the free pre-Lie
algebra on one generator \cite{CL}.
\subsection{Tree-like structures and variations on a theme}
\subsubsection*{Tree-like structures}
From the Hopf algebra $\mathcal{H},$ defined in the previous
subsection, several generalizations can be constructed: Hopf
algebras of decorated trees, of planar trees, etc. This can be
phrased most elegantly from a general point of view in terms of
''tree-like structures``, as for example introduced by Turaev in
\cite{Turaev}: Consider the category of rooted trees and embeddings
(an embedding $\tau'\rightarrow \tau$ is an isomorphism from $\tau'$
to a subtree of $\tau$). A rooted tree-structure is then defined to
be a contravariant functor from this category to the category of
sets. For example, decorated (labelled) trees can be described by
the functor $\phi$ which maps a tree onto a certain set its vertices
and/or edges are decorated with. Being contravariant, $\phi$ maps
embeddings of trees to the respective restrictions of decorations.
Similarly, a planar structure is provided by a functor $\phi$
mapping a tree to the set of its topological embeddings into the
real plane modulo orientation-preserving homeomorphisms of $\R^2$
onto itself. Now let $\phi$ be a rooted tree-structure. A rooted
$\phi$-tree is a pair $(\tau,s)$ where $\tau$ is a tree and $s$ is
an element of $\phi(\tau).$ The notions of isomorphisms and subtrees
of rooted $\phi$-trees are immediate.
\subsubsection*{Generalizations of $\mathcal{H}$}
Using this convenient framework, we have immediately other Hopf
algebras at hand: Let $S$ be a set. The Hopf algebra
$\mathcal{H}(S)$ is defined as in the previous subsection,
replacing the word tree by $S$-decorated tree (for our purposes,
we only decorate vertices, not edges). Similarly,
$\mathcal{H}_{pl}$ is the (noncommutative) Hopf algebra of planar
rooted trees. In particular, for these Hopf algebras, the proofs
of the coassociativity of $\Delta$ are verbatim the same. The
planar Hopf algebra and its decorated versions $H_{pl}(S)$ were
extensively studied by Foissy \cite{FoissyBSM1}. He showed that
they are self-dual and constructed isomorphisms to several other
Hopf algebras on trees that have appeared in the literature.
\subsubsection*{The Hopf algebra of Feynman graphs}
While rooted trees describe nested divergences in an obvious
manner, the resolution of \emph{overlapping} divergences into
trees requires some care \cite{Zimmermann,Kreimer6,FGB}. This
problem exists only in momentum space. By basing a Hopf algebra
directly on Feynman graphs instead of trees, these issues can be
avoided \cite{Kreimer6,CK2}. As an algebra, let $\mathcal{H}_{CK}$
be the free commutative algebra on 1PI Feynman graphs (of a given
theory; the case of a non-scalar theory requires to take form
factors (external structures) into account which we avoid here).
The empty graph serves as a unit $\One.$ In the following, a
product of graphs is identified with the disjoint union of these
graphs. On a graph, a coproduct is given \cite{CK2} by
\begin{equation*}
\Delta(\Gamma)=\One\otimes\Gamma+\Gamma\otimes\One+\sum_{\gamma\subsetneq
\Gamma} \gamma \otimes \Gamma/\gamma
\end{equation*}
where the sum is over all 1PI superficially divergent proper
subgraphs $\gamma$ of $\Gamma.$ A few examples are given in
\cite{CK2}. \\\\
Still, thanks to the universal property mentioned at the end of
subsection \ref{ss:Hochschild}, Hopf algebras of rooted trees serve
as an excellent role model for various questions and, moreover,
yield most interesting links to different branches of mathematics
\cite{CM,ggl}. In the present paper, we will be mainly concerned
with Hopf algebras of trees. In many cases, it is only a matter of
notation to translate these results into the Hopf algebra of Feynman
graphs,
easily achieved by the practitioner of QFT \cite{BroadKreimer2}. \\\\
In view of the preceding paragraphs, the reader might wish to try to
describe $\mathcal{H}_{CK}$ as a Hopf algebra of suitable tree-like
structures, using the results of \cite{Kreimer6}.
\subsection{Hochschild cohomology of bialgebras}\label{ss:Hochschild}
\subsubsection*{Definition}
Let $A$ be a bialgebra. We consider linear maps $L: A\rightarrow A^{\otimes n}$ as $n$-cochains and define a coboundary operator $b$ by
\begin{equation}
\label{Hscb} bL := (id\otimes L)\Delta+\sum_{i=1}^n(-1)^i\Delta_i
L+(-1)^{n+1}L\otimes \One
\end{equation}
where $\Delta$ denotes the coproduct and $\Delta_i$ the coproduct
$\Delta$ applied to the $i$-th factor in $A^{\otimes n}$. The map
$L\otimes\One$ is given by $x\mapsto L(x)\otimes\One.$ It is
essentially due to the coassociativity of $\Delta$ that $b$ squares
to zero, which gives rise to a cochain complex $(C,b).$ Clearly
$(C,b)$ captures only information about the coalgebra structure of
$A.$
The cohomology of $(C,b),$ denoted $HH^\bullet_\epsilon(A),$ is
easily seen to be the dual ($A$ considered as a bi\emph{co}module
rather than a bimodule over itself) notion of the Hochschild
cohomology of algebras. Note that the right bicomodule action is
 here $(id\otimes
\epsilon)\Delta$ which explains the last summand in (\ref{Hscb})
and the subscript in $HH^\bullet_\epsilon.$
\subsubsection*{The role of $HH_\epsilon^1(\mathcal{H})$}
For $n=1,$ the cocycle condition $bL=0$ reduces to, for $L:
A\rightarrow A,$
\begin{equation}\label{eq:Hscb1}
\Delta L = (id\otimes L) \Delta+L\otimes\One.
\end{equation}
Sometimes the following equivalent statement, using the map
$\tilde\Delta,$ is more convenient:
\begin{equation}\label{eq:Hscb2}
\tilde\Delta L = (id\otimes L)\tilde\Delta+id\otimes L(\One).
\end{equation}
Let us now try to understand the space $HH_\epsilon^1(\mathcal{H})$
of ''outer coderivations on $\mathcal{H}$.``
 We first describe the 0-coboundaries (''inner coderivations``). They
are of the form
\begin{equation*}
L(\tau) = \sum \alpha_{\tau''}\tau'-\alpha_{\tau}\One
\end{equation*}
in Sweedler's notation, where $\alpha_\tau$ is an element of $k$ for
each forest $\tau.$ For example, $L: \tau\mapsto \sum \tau'-\One$ is
a
0-coboundary. Note that $\One$ is in the kernel of any 0-coboundary.\\\\
It is a crucial fact that the grafting operator $B_+$, introduced in
 subsection \ref{ss:trees}, is a 1-cocycle \cite{CK}:
 \begin{pro}
\begin{equation}\label{eq:Hscb3}
\Delta B_+ = (id\otimes B_+)\Delta+B_+\otimes\One.
\end{equation}
\begin{equation}\label{eq:Hscb4}
\tilde\Delta B_+ = (id\otimes B_+)\tilde\Delta+id\otimes\bullet.
\end{equation}
\end{pro}
\emph{Idea of proof: } When looking at equation (\ref{eq:Hscb4}),
the statement is rather immediate: Let $\tau$ be a forest. The
first summand at the right side of (\ref{eq:Hscb4}) refers to cuts
of $B_+(\tau)$ which affect at most all but one of the edges
connecting the new root of $B_+(\tau)$ to the roots of $\tau,$
while the second summand takes care of the cut which completely
separates the root of $B_+(\tau)$
from all its children. \hfill $\Box$ \\\\
Since $B_+$ is a homogeneous linear endomorphism of degree 1, it is
not a 0-coboundary -- note that the coboundaries have no chance to
increase the degree. Thus $B_+$ is a generator (among
others) of $HH_\epsilon^1(\mathcal{H}).$\\\\
When looking for other generators $L$ of
$HH_\epsilon^1(\mathcal{H}),$ the cocycle conditions
(\ref{eq:Hscb1},\ref{eq:Hscb2}) immediately yield the requirement
that $L(\One)$ be a primitive element (and zero if $L$ is exact).
While $\bullet$ is up to scalar factors obviously the only
primitive element in degree 1, there are plenty of primitives in
higher degrees. For example,
\begin{equation}\label{eq:primi}
\bullet\bullet - 2 \oline
\end{equation}
is a primitive element in degree 2. Foissy \cite{FoissyBSM1}
showed that $L\mapsto L(\One)$ is a surjective map
$HH^1_\epsilon(\mathcal{H})\rightarrow $Prim$(\mathcal{H})$ onto
the set of primitive elements of $\mathcal{H}.$ In the case of
Hopf algebras of decorated rooted trees $\mathcal{H}(S)$ obviously
any element $s\in S$ yields a homogeneous cocycle of degree 1
denoted $B_+^s$ which, applied to a
forest, connects its roots to a new root decorated by $s.$\\\\
It should be clear that each 1PI Feynman graph which is free of
subdivergences is a primitive element of $\mathcal{H}_{CK}.$ In
general, there are primitive elements in higher degrees too, for
example, cf.~(\ref{eq:primi}), the linear combination
\begin{equation*}
\phiright\phiright-2\phidouble
\end{equation*}
in $\phi^3$ theory in six dimensions.
\subsubsection*{Universal property}
The category of objects $(A,L)$ consisting of a commutative
bialgebra $A$ and a Hochschild 1-cocycle $L$ on $A$ with morphisms
bialgebra morphisms commuting with the cocycles has the initial
object $(\mathcal{H},B_+)$. This is a result of \cite{CK}. Indeed,
let $(A,L)$ be such a pair. The map $\rho: \mathcal{H}\rightarrow A$
is simply defined by $\rho(\One)=\One$ and pushing forward along
$B_+$ (and $L$) and the multiplication. The fact that $\rho$ is a
morphism of coalgebras is an easy consequence of (\ref{eq:Hscb3}).\\\\
Also it was shown in \cite{BK} that, conversely, the coproduct
$\Delta$ of $\mathcal{H}$ is determined if one requires the map
$B_+$ to be a 1-cocycle. This may serve to find different
presentations of $\mathcal{H}.$\\\\
For any $\mathcal{H}$-bicomodule $B$, the higher Hochschild
cohomology $HH^n(\mathcal{H},B),$ $n\ge 2,$ is trivial
\cite{FoissyBSM1}, thus in particular
$HH_\epsilon^n(\mathcal{H})=0.$
\subsection{Finiteness and locality from the Hopf
algebra}\label{ss:locality}
We have now accumulated enough algebraic notions to come back to
the original physical application already sketched in subsection
\ref{ss:motivation}. Given a specific quantum field theory, Hopf
algebras $\mathcal{H}(S)$ and $\mathcal{H}_{CK}$ are determined by
its perturbative expansion into Feynman graphs. We denote this
Hopf algebra generically by $\mathcal{H}.$ Every divergent graph
$\gamma$ without subdivergences determines a Hochschild 1-cocycle
$B_+^\gamma,$ and any relevant tree or graph is in the range of a
1-cocycle of this kind. This ensures that any relevant term in the
perturbative expansion is in the image of a Hochschild
one-cocycle. This allows to prove locality: the all important
Bogoliubov $\bar{R}$ operation on a character $\phi$, \be \phi\to
\bar{R}[\phi]=m(S_R^\phi\otimes \phi )({\rm id}\otimes P)\Delta,
\ee $P$ the projector in the augmentation ideal and $\phi$ some
Feynman rules has the property that it only requires an overall
Taylor subtraction at a fixed renormalization point to render it
finite - implying that its divergences  will not depend on
logarithms of kinematical variables.

\subsubsection*{Momentum space}
The next step is to choose a target algebra $V$ and regularized
Feynman rules $\phi: \mathcal{H}\rightarrow V,$ and a
renormalization scheme $R: V\rightarrow V.$ The map $\phi$ is
supposed to be a (unital) algebra homomorphism. We stick to the
example $(V=\C[\epsilon^{-1},\epsilon]],\phi)$ of dimensional
regularization as in subsection \ref{ss:motivation}, but stress once
more that the reader can find suitable generalizations in the
literature \cite{EFGK2}. The minimal subtraction scheme where $R$ is
the projector onto the proper pole part is only one of many choices
one can make. However, in any case we require $R$ to preserve the UV
divergent structure (i.~e.~ the pole part) and to satisfy the
Rota-Baxter equation
\begin{equation}\label{eq:rb}
R(xy)+R(x)R(y)=R(xR(y))+R(R(x)y).
\end{equation}
It is easy to check that the minimal subtraction scheme satisfies
(\ref{eq:rb}). The Rota-Baxter equation is the algebraic key to the
link between renormalization and Birkhoff decomposition, see for
example \cite{CK2,EFGK1,EFGK2}. It also guarantees that the
renormalized Feynman rules are again an algebra homomorphism
\cite{Kreimer4} as are the unrenormalized rules $\phi.$ Now the
twisted antipode is defined by
\begin{equation}\label{eq:twisted}
S_R^\phi(\tau) = -R(S_R^\phi\star_P \phi)(\tau)\mbox{ for
}\tau\not\in\mathcal{H}_0,\quad S_R^\phi(\One)=1,
\end{equation}
equivalently, in Sweedler's notation
\begin{equation*}
S_R^\phi(\tau) = -R\left(\phi(\tau)+ \tilde\sum
S_R^\phi(\tau')\phi(\tau'')\right)\mbox{ for
}\tau\not\in\mathcal{H}_0,\quad S_R^\phi(\One)=1
\end{equation*}
 where the term ''twisted
antipode`` should be justified by a glance at the recursive
expression (\ref{eq:antipode}) for the regular antipode. The map
$S_R^\phi,$ as can be inferred from the example in Figures 1--3,
yields the counterterm for $\phi.$ The complete renormalized
evaluation function is then given by
\begin{equation}\label{eq:phir}
\phi_R = S_R^\phi\star \phi.
\end{equation}
One can find a non-recursive description of $\phi_R$
\cite{Kreimer,CK} which shows the equivalence with Zimmermann's
forest formula \cite{Zimmermann}. \\\\
\subsubsection*{Example}
In order to understand the twisted antipode, we come back to the
example of subsection \ref{ss:motivation}. On the relevant trees,
the coproduct acts as follows:
\begin{eqnarray}\label{eq:nutshelldelta}
\Delta\left(\ocgammas\right) &=&
\One\otimes\ocgammas+\ocgammas\otimes\One+\nonumber\\
&&+ \bullet_{\gamma_2}\otimes\olinegammaonethree+
\bullet_{\gamma_3}\otimes\olinegammaonetwo+
\bullet_{\gamma_2}\bullet_{\gamma_3}\otimes\bullet_{\gamma_1},\\
\Delta(\bullet_{\gamma_i}) &=&
\One\otimes\bullet_{\gamma_i}+\bullet_{\gamma_i}\otimes
\One,\nonumber
\end{eqnarray}
According to (\ref{eq:twisted}) and (\ref{eq:phir}), the algorithm
for $\phi_R$ consists of the following steps:
\begin{enumerate}
\item[($F$)] Apply the coproduct $\Delta$ to the tree under
consideration \item[($C_n$)] apply the map $(id\otimes
P)\Delta\otimes id^{\otimes n}$ (for $n=1\ldots$) until each
summand is of the form $\One\otimes \ldots.$ \item[($M$)] apply
$\phi^{\otimes n}$ to go into $V^{\otimes n}.$ As
$\phi(\One)=S_R^\phi(\One)=1,$ the first factor $\One$ of each
term is mapped to 1. \item[($C'_n$)] (for $n=\ldots 1$) apply the
map $-Rm\otimes id^{\otimes n}$ until we end up in $V^{\otimes 2}$
\item[($F'$)] apply the map $m$ to get into $V.$
\end{enumerate}
For the tree $\ocgammas$ this algorithm is performed in Figures
1--3.
\begin{figure}\label{fig:one}
\please\begin{diagram}
\mathcal{H} &&  \ocgammas\\
\dTo>{\Delta} &&  \\
\stackrel{\phi^{\otimes 2}}{\leftarrow}\mathcal{H}^{\otimes 2} &&
\ocgammas\otimes
\One+\One\otimes\ocgammas+\bullet_{\gamma_2}\otimes\olinegammaonethree+
\bullet_{\gamma_3}\otimes\olinegammaonetwo+
\bullet_{\gamma_2}\bullet_{\gamma_3}\otimes\bullet_{\gamma_1}\\
\dTo>{(id\otimes P)\Delta\otimes id}&&\\
\stackrel{\phi^{\otimes 3}}{\leftarrow}\mathcal{H}^{\otimes 3} &&
\left(\One\otimes\ocgammas+\bullet_{\gamma_2}\otimes\olinegammaonethree+
\bullet_{\gamma_3}\otimes\olinegammaonetwo+
\bullet_{\gamma_2}\bullet_{\gamma_3}\otimes\bullet_{\gamma_1}\right)\otimes \One\\
&& +(\One\otimes\bullet_{\gamma_2}\bullet_{\gamma_3}+
\bullet_{\gamma_2}\otimes\bullet_{\gamma_3}+\bullet_{\gamma_3}\otimes\bullet_{\gamma_2})\otimes \bullet_{\gamma_1}\\
\dTo>{(id\otimes P)\Delta\otimes id^{\otimes 2}}&& +\One\otimes
\bullet_{\gamma_2}\otimes
\olinegammaonethree+\One\otimes\bullet_{\gamma_3}\otimes\olinegammaonetwo\\
&& +1\mbox{ summand done (was already of the form
}\One\otimes\ldots)
\\
\stackrel{\phi^{\otimes 4}}{\leftarrow}\mathcal{H}^{\otimes 4} &&
\One\otimes\bullet_{\gamma_2}\otimes \olinegammaonethree\otimes \One
+\One\otimes\bullet_{\gamma_3}\otimes \olinegammaonetwo\otimes
\One+\One\otimes\bullet_{\gamma_2}\bullet_{\gamma_3}\otimes\bullet_{\gamma_1}\otimes
\One\\
&&
+\bullet_{\gamma_2}\otimes\bullet_{\gamma_3}\otimes\bullet_{\gamma_1}\otimes
\One+\bullet_{\gamma_3}\otimes\bullet_{\gamma_2}\otimes\bullet_{\gamma_1}\otimes
\One\\
&&+
\One\otimes\bullet_{\gamma_2}\otimes\bullet_{\gamma_3}\otimes\bullet_{\gamma_1}+
\One\otimes\bullet_{\gamma_3}\otimes\bullet_{\gamma_2}\otimes\bullet_{\gamma_1}\\
\dTo>{(id\otimes P)\Delta\otimes id^{\otimes 3}}&&+5\mbox{ summands done}\\
\stackrel{\phi^{\otimes 5}}{\leftarrow}\mathcal{H}^{\otimes 5} &&
\One\otimes
\bullet_{\gamma_2}\otimes\bullet_{\gamma_3}\otimes\bullet_{\gamma_1}\otimes
\One+\One\otimes
\bullet_{\gamma_3}\otimes\bullet_{\gamma_2}\otimes\bullet_{\gamma_1}\otimes
\One\\ &&+10\mbox{ summands done}
\end{diagram}
\caption{First part of the calculation of $\phi_R.$ Apply $\Delta$
and then $(id\otimes P)\Delta\otimes id^{\otimes n}$ until each
summand is of the form $\One\otimes \ldots.$}
\end{figure}
\begin{figure}\label{fig:two}
\please\begin{diagram} \stackrel{\phi^{\otimes 5}}{\rightarrow}
V^{\otimes 5} && \ldots \\
\dTo>{-Rm\otimes id^3}&&\\
\stackrel{\phi^{\otimes 4}}{\rightarrow}V^{\otimes 4}&&
-R\phi(\bullet_{\gamma_2})\otimes\phi(\bullet_{\gamma_3})\otimes\phi(\bullet_{\gamma_1})\otimes
1-R\phi(\bullet_{\gamma_3})\otimes\phi(\bullet_{\gamma_2})\otimes\phi(\bullet_{\gamma_1})\otimes 1\\
\dTo>{-Rm\otimes id^2}&&+10\mbox{ summands
pending}\\
\stackrel{\phi^{\otimes 3}}{\rightarrow}V^{\otimes 3}&&
-R\phi(\bullet_{\gamma_2})\otimes
\phi\left(\olinegammaonethree\right)\otimes 1
-R\phi(\bullet_{\gamma_3})\otimes
\phi\left(\olinegammaonetwo\right)\otimes 1\\
&&
-R\phi(\bullet_{\gamma_2}\bullet_{\gamma_3})\otimes\phi(\bullet_{\gamma_1})\otimes
1 + R(
R\phi(\bullet_{\gamma_2})\phi(\bullet_{\gamma_3}))\otimes\phi(\bullet_{\gamma_1})\otimes
 1\\
&&+ R(
R\phi(\bullet_{\gamma_3})\phi(\bullet_{\gamma_2}))\otimes\phi(\bullet_{\gamma_1})\otimes
1-R\phi(\bullet_{\gamma_2})\otimes\phi(\bullet_{\gamma_3})\otimes\phi(\bullet_{\gamma_1})\\
\dTo>{-Rm\otimes id}&& -
 R\phi(\bullet_{\gamma_3})\otimes\phi(\bullet_{\gamma_2})\otimes\phi(\bullet_{\gamma_1})\\
&& +5\mbox{ summands pending}\\
\stackrel{\phi^{\otimes 2}}{\rightarrow}V^{\otimes 2}&&
R\left(R\phi(\bullet_{\gamma_2})\phi\left(\olinegammaonethree\right)\right)\otimes
1
+R\left(R\phi(\bullet_{\gamma_3})\phi\left(\olinegammaonetwo\right)\right)\otimes
1\\
&&
+R\left(R\phi(\bullet_{\gamma_2}\bullet_{\gamma_3})\phi(\bullet_{\gamma_1})\right)\otimes
1 -R\left(R(
R\phi(\bullet_{\gamma_2})\phi(\bullet_{\gamma_3}))\phi(\bullet_{\gamma_1})\right)\otimes
1\\
&&-R\left( R(
R\phi(\bullet_{\gamma_3})\phi(\bullet_{\gamma_2}))\phi(\bullet_{\gamma_1})\right)\otimes
1+R\left(R\phi(\bullet_{\gamma_2})\phi(\bullet_{\gamma_3})\right)\otimes\phi(\bullet_{\gamma_1})\\
\dTo>{m} && +R\left(R\phi(\bullet_{\gamma_3})\phi(\bullet_{\gamma_2})\right)\otimes\phi(\bullet_{\gamma_1})-R\phi\left(\ocgammas\right)\otimes 1\\
&&-R\phi(\bullet_{\gamma_2}\bullet_{\gamma_3})\otimes \phi(\bullet_{\gamma_1})-R\phi(\bullet_{\gamma_2})\otimes \phi\left(\olinegammaonethree\right)\\
&&-R\phi(\bullet_{\gamma_3})\otimes \phi\left(\olinegammaonetwo\right)+1\mbox{ summand pending}\\
\end{diagram}
\caption{Second part of the calculation of $\phi_R$. Apply
$\phi^{\otimes n}$ and then $- Rm\otimes id^{\otimes n}$ until
arrival in $V^{\otimes 2}.$ Then apply $m$ to get into $V.$}
\end{figure}
\begin{figure}\label{fig:three}
\please\begin{diagram}
\dTo>{m}&&\\
V && - R\phi\left(\ocgammas\right)+ R\left(
R\phi(\bullet_{\gamma_2})\phi\left(\olinegammaonethree\right)\right)
+ R\left( R\phi(\bullet_{\gamma_3})\phi\left(\olinegammaonetwo\right)\right)\\
&& + R\left(
R\phi(\bullet_{\gamma_2}\bullet_{\gamma_3})\phi(\bullet_{\gamma_1})\right)-
 R\left( R(
R\phi(\bullet_{\gamma_2})\phi(\bullet_{\gamma_3}))\phi(\bullet_{\gamma_1})\right)
\\
&&- R\left( R(
R\phi(\bullet_{\gamma_3})\phi(\bullet_{\gamma_2}))\phi(\bullet_{\gamma_1})\right)+\phi\left(\ocgammas\right)-
R\phi(
\bullet_{\gamma_2})\phi\left(\olinegammaonethree\right)\\
&&- R\phi( \bullet_{\gamma_3})\phi\left(\olinegammaonetwo\right)-
R\phi(\bullet_{\gamma_2}\bullet_{\gamma_3})
\phi(\bullet_{\gamma_1})\\
&&+ R\left(
R\phi(\bullet_{\gamma_2})\phi(\bullet_{\gamma_3})\right)\phi(\bullet_{\gamma_1})
+ R\left(R\phi(\bullet_{\gamma_3})\phi(\bullet_{\gamma_2})\right)\phi(\bullet_{\gamma_1})\\
\end{diagram}
\caption{Third part of the calculation of $\phi_R$. The reader
should compare the result with (\ref{eq:Bogo}). Using the fact
that $S_R^\phi$ is an algebra homomorphism (if $R$ is a
Rota-Baxter map), the last step (C3) in Figure 1 and the first
step (C3') in Figure 2 could have been avoided.}
\end{figure}
While in our simple example of only two disjoint subdivergences, the
Bogoliubov recursion could have been performed by hand without using
the Hopf algebra, when going to higher loop orders, the Hopf algebra
approach provides significant computational advantage
\cite{BroadKreimer2,BroadKreimer4}.
\subsubsection*{Locality of counterterms from Hochschild cohomology}
Moreover, the Hopf algebra can be used to give a direct proof of
finiteness of renormalization and locality of counterterms from a
purely algebraic point of view. For a simple toy model, this has
been done in a recent paper \cite{Kreimer2}. The basic observation
is that the fact that every relevant tree or graph is in the range
of a homogeneous Hochschild 1-cocycle of degree 1 allows for easy
and clean inductive proofs of various statements for arbitrary
loop number. This also holds on the level of graphs: the sum over
all primitive graphs of given loop order $n$ defines a 1-cocycle
$B_+^n$ such that every graph is generated in the range of these
1-cocycles. This allows, as observed in \cite{annals,dennisfest},
to
prove locality in general. \\\\
Indeed, let $B_+^n$ be a 1-cocycle, and let $\mu_+$ be the measure
defined by the $n$-loop integrand of $B_+^n(\One)$. Let
$\phi(B_+^n(X))$ be a Feynman amplitude defined by insertion of a
collection of subdivergences $X$ into those $n$-loop primitive
graphs. We write $\phi(B_+^n(X))=\int \phi(X)d\mu_+,$ emphasizing
that subgraphs become subintegrals under the Feynman
rules. \\\\
Recall that $P$ denotes the projection onto the augmentation
ideal. Since $P(\One)=0$, $PB_+^n=B_+^n,$ we can write \be m
(S_R^\phi\otimes\phi P)\Delta(B_+^n(X))=\int
S_R^\phi\star\phi(X)d\mu_+.\ee This proves locality in a
straightforward manner by induction over the augmentation degree,
i.~e.\ using the coradical filtration of the Hopf algebra.
\subsubsection*{Coordinate space}
The language of rooted trees is especially suited for describing
renormalization in coordinate space \cite{BK}. A particularly
appealing approach to coordinate space renormalization is the work
of Epstein and Glaser \cite{EG} (see also \cite{PS,BF}) who,
starting from ideas of Bogoliubov \cite{BS} and others, extracted
a set of axioms for time-ordered product and constructed such
time-ordered products in terms of rigorous functional analysis.
The result is completely equivalent to momentum space
renormalization but has conceptual (albeit not computational)
advantages. It is no surprise that the Hopf algebra picture fits
equally nice into this framework \cite{BK}, if one takes into
account the specific features of Epstein-Glaser renormalization
such as the absence of overlapping divergences and regularization
parameters. In view of highly interesting mathematical
ramifications such as a possible analogy to the Fulton-MacPherson
compactification of configuration spaces \cite{FM,MSS}, it seems
most appropriate to attack this problem using trees \cite{BK}
rather than coordinate space Feynman diagrams. \\\\
If there are no subdivergences, Epstein-Glaser renormalization
amounts to a Taylor subtraction on test functions: Let $^0t$ be a
distribution on some $\R^d-\{0\}$ with singularity at $0,$ for
example $^0t = x^{-(d+1)}.$ In order to extend $^0t$ onto all of
$\R^d,$ consider
\begin{equation}\label{eq:egsubtr}
t: f\mapsto \left.^0t\right.\left(f-\sum_{|\alpha|\le
\rho}w_\alpha
\partial^\alpha f(0)\right)
\end{equation}
where $f$ is a test function and the $w_\alpha$ are auxiliary test
functions with $\partial^\beta w_\alpha(0)=\delta_\alpha^\beta.$
If $\rho$ is large enough with respect to the degree of divergence
of $^0t$ at $0,$ the modified distribution $t$ is defined on all
of $\R^d.$ It is natural to consider the first summand in
(\ref{eq:egsubtr}) as the unrenormalized contribution and the
second
summand as the counterterm.\\\\
Epstein and Glaser describe how to take care of distributions which
may have an overall divergence and subdivergences, i.~e.~
distributions which are not only singular on the thin diagonal
$\{x_1=\ldots=x_n\}$ of some $M^n$ but on the fat diagonal
$\{x_i=x_j\mbox{ for some }i,j\}.$ The algorithm, with the above
identification of unrenormalized part and counterterm, is
structurally very similar to the Bogoliubov recursion and can thus
be described by a twisted antipode \cite{BK}.\\\\
Using some techniques of \cite{BK}, notably the ''cut product``
$\odot$ of certain linear endomorphisms on $\mathcal{H}$ as a
replacement for the convolution product, one can construct the map
$R$ as an algebra endomorphism on a Hopf algebra of trees with
decorated vertices $\mathcal{H}(\{\bullet,\ast\})$ and consider the
''twisted antipode`` defined by $S_R=-R(S_R\odot_P id),$ and the
renormalization map $S_R\odot id.$ Starting from the tree $\oc,$ the
map $S_R\odot id$ yields
\begin{equation*}
-\octop+2\oclefttop-\ocleftrighttop+\oc-2\ocleft+\ocleftright
\end{equation*}
which should be compared to the last line in Figure 3 -- vertices
of type $\bullet$ mark unrenormalized contributions, vertices of
type $\star$ the corresponding counterterms. These trees are then
mapped into an appropriate space of operator-valued distributions:
the above example describes terms needed for the renormalization
of the fourth order time-ordered product. Using this somewhat
modified approach, where the combinatorics happen entirely in the
Hopf algebra (as opposed to between the Hopf algebra and the
target ring), checking locality simply amounts to calculating the
commutator of $S_R\odot id$ and $B_{+\bullet}:$
\begin{equation*}
(S_R\odot id)B_{+\bullet}=(id-R)B_{+\bullet}(S_R\odot id).
\end{equation*}
Thus once the subdivergences are taken care of, it suffices to
subtract the superficial divergence.
\section{Hopf subalgebras and Dyson-Schwinger equations}
Hopf subalgebras of the Hopf algebras of (decorated) rooted trees
or Feynman graphs are in close relationship with Dyson-Schwinger
equations. Indeed, any Dyson-Schwinger equation (to be defined
below) gives rise to a Hopf subalgebra. This is a statement about
self-similarity: a 1-cocycle like $B_+$ ensures that a product of
trees is mapped to a tree, and this is a rather general
phenomenon: the Green functions appear as functionals of
themselves, the functionals being provided by the Dyson skeleton
graphs which appear
as the integral kernel $\phi(B_+^n(\One))$. \\\\
It will turn out in \tref{thm:dsesub} that all Hopf subalgebras
coming from a reasonably general class of Dyson-Schwinger equations
are in fact isomorphic.
\subsection{Hopf subalgebras of decorated rooted trees}
For simplicity, we start our considerations in the Hopf algebra
$\mathcal{H}$ of undecorated rooted trees. A full classification of
their Hopf subalgebras is far beyond reach. However, we give a few
examples the last of which will be directly related to
Dyson-Schwinger equations.
\subsubsection*{Bounded fertility, finite parts, primitive elements}
For $n\in\N$ let $\mathcal{H}_n$ be the subalgebra of
$\mathcal{H}$ generated by trees whose vertices have fertility
bounded from above by $n.$ A glance at the definition of the
coproduct (\ref{cuts}) suffices to see that $\mathcal{H}_n$ is a
Hopf subalgebra of $\mathcal{H}.$ In particular, the Hopf algebra
$\mathcal{H}_1$ with one generator in each degree is known as the
Hopf algebra of \emph{ladders}. It is closely related to iterated
integrals \cite{Chen,Kreimer4}.\\\\
Similarly, the free commutative algebra generated by trees of
degree $\le n$ forms a Hopf subalgebra for any $n$ since the
coproduct respects the grading. Another example where there is
nothing to check are subalgebras generated by an arbitrary
collection of primitive elements of $\mathcal{H}.$
\subsubsection*{The Connes-Moscovici Hopf subalgebra}
A less trivial example of a Hopf subalgebra of $\mathcal{H}$ arose
in the work of Connes and Moscovici on local index formulas for
transversally hypoelliptic operators on foliations
\cite{CM,CK,CM2}. In the case of a foliation of codimension 1, the
relevant Hopf algebra $\mathcal{H}_{T}$ is defined by the
generators $X,Y,$ $\delta_n$ for $n\in\N,$ the relations
\begin{equation*}
[X,Y]=-X,\quad
[X,\delta_n]=\delta_{n+1},\quad[Y,\delta_n]=n\delta_n,\quad
[\delta_n,\delta_m]=0,
\end{equation*}
and the coproduct
\begin{equation*}
\Delta(X)=X\otimes\One+\One\otimes X+\delta_1,\quad
\Delta(Y)=Y\otimes\One+\One\otimes Y, \quad
\Delta(\delta_1)=\delta_1\otimes \One+\One\otimes\delta_1.
\end{equation*}
Note that the relations above and the requirement that $\Delta$ be
an algebra homomorphism determine $\Delta$ on the generators
$\delta_n$ for $n\ge 2$ as well. Let $N$ be the linear operator,
called \emph{natural growth operator}, on $\mathcal{H},$ defined
on a tree $\tau$ by adding a branch to each vertex of $\tau$ and
summing up the resulting trees, extended as a derivation onto all
of $\mathcal{H}.$ For example,
\begin{eqnarray}\label{eq:natgrowth}
N(\One)&=&\bullet,\nonumber\\
N^2(\One)&=&\oline,\nonumber\\
N^3(\One)&=&\oc+\olongline,\\
N^4(\One)&=&\othree+3\occ+\ofork+\olonglongline.\nonumber
\end{eqnarray}
Now identifying $\delta_1$ with $\bullet,$ and generally
$\delta_n$ with $N^n(\One),$ the commutative Hopf subalgebra of
$\mathcal{H}_T$ generated by the $\delta_n$ can be embedded into
$\mathcal{H}$ \cite{CK}. The resulting Hopf subalgebra is denoted
$\mathcal{H}_{CM}.$ For example,
\begin{eqnarray*}
\tilde \Delta(\delta_1) &=& 0,\\
\tilde \Delta(\delta_2) &=& \delta_1\otimes \delta_1,\\
\tilde \Delta(\delta_3) &=&
3\delta_1\otimes\delta_2+(\delta_2+\delta_1^2)\otimes\delta_1.
\end{eqnarray*}
The $\delta_n$ can be specified in a non-recursive manner:
\begin{equation*}
\delta_n=\sum_{\tau\in\mathcal{T}_n} c_\tau \tau.
\end{equation*}
Here $\mathcal{T}_n$ is the set of trees of weight $n.$ The integers
$c_\tau$, called \emph{Connes-Moscovici weights}, have been computed
in \cite{Kreimer4,FoissyBSM1} using the tree factorial
\begin{equation*}
c_\tau = \frac{n!}{\tau!\,\mbox{Sym}(\tau)}
\end{equation*}
where $\mbox{Sym}(\tau)$ is the symmetry factor (rank of the group
of symmetries) of $\tau.$
\subsubsection*{A quadratic Dyson-Schwinger equation}
Now we turn to the study of another source of Hopf subalgebras,
the combinatorial Dyson-Schwinger equations. As a first example,
we consider the equation
\begin{equation}\label{eq:quadrdse}
X= \One+\alpha B_+(X^2)
\end{equation}
in $\mathcal{H}[[\alpha]].$
Using the ansatz
\begin{equation*}
X = \sum_{n=0}^\infty \alpha^n c_n
\end{equation*}
one easily finds $c_0=\One$ and
\begin{equation}\label{eq:reccn}
c_{n+1}= \sum_{k=0}^n B_+(c_k c_{n-k})
\end{equation}
which determine $X$ by induction. The first couple of $c_n$ are
easily calculated:
\begin{eqnarray*}
c_0 & = & \One,\\
c_1 & = & \bullet,\\
c_2 & = & 2 \oline\\
c_3 & = & \oc + 4 \olongline \\
c_4 & = & 4\occ+2\ofork+8\olonglongline
\end{eqnarray*}
We observe that $c_n$ is a weighted sum of trees with vertex
fertility bounded by 2 -- this is due to the square of $X$ in the
Dyson-Schwinger equation (\ref{eq:quadrdse}). The reader should
compare this to the Connes-Moscovici trees (\ref{eq:natgrowth})
discussed in the previous subsection. The recursive nature of
(\ref{eq:quadrdse}) makes one suspect that the $c_n$ generate a Hopf
subalgebra of $\mathcal{H}.$ Indeed, for each $n\ge 0$ and $k\le n$
there is a polynomial $P^n_k$ in the $c_l$ for $l\le n$ such that
\begin{equation}\label{eq:recdeltacn}
\Delta c_{n} = \sum_{k=0}^{n} P^n_k \otimes c_k.
\end{equation}
They are inductively determined by
\begin{equation}\label{eq:recpnk}
P^{n+1}_{k+1}= \sum_{l=0}^{n-k} P_0^{l}P_k^{n-l}
\end{equation}
and $P^{n+1}_0 = c_{n+1}.$
For a proof of this statement, see the more general
\tref{thm:dsesub} in the next subsection.
For the moment, we merely display the first $P^n_k$ in an upper
triangular matrix where columns are indexed by $n=0\ldots 5$ and
rows by $k=0\ldots n.$
\begin{equation*}
\left[ \begin{array}{cccccccc}
\One    & c_1       & c_2       & c_3           & c_4           & c_5        \\
       & \One      & 2c_1      & 2c_2+c_1^2        & 2c_3+2c_1c_2      & 2c_4+2c_1c_3+c_2^2 \\
       &      & \One      & 3c_1          & 3c_2+3c_1^2       & 6c_1c_2 + c_1^3+3c_3  \\
       &      &      & \One          & 4c_1          & 6c_1^2+4c_2  \\
   &      &      &          & \One          & 5c_1     \\
   &      &      &          &          & \One
\end{array} \right]
\end{equation*}
The coefficients are basically multinomial coefficients as will
become clear in the next subsection.
\subsection{Combinatorial Dyson-Schwinger equations}
Let $A$ be any connected graded Hopf algebra which is free or free
commutative as an algebra, and $(B_+^{d_n})_{n\in\N}$ a collection
of Hochschild 1-cocycles on it (not necessarily pairwise
distinct). The most general Dyson-Schwinger equation we wish to
consider here is
\begin{equation}\label{eq:genDSE}
X=\One+\sum_{n=1}^\infty \alpha^n w_n  B_+^{d_n}(X^{n+1})
\end{equation}
in $A[[\alpha]].$ The parameter $\alpha$ plays the role of a
coupling constant. The $w_n$ are scalars in $k.$ Again we
decompose the solution
\begin{equation*}
X=\sum_{n=0}^\infty \alpha^n c_n
\end{equation*}
with $c_n\in A.$
\begin{lem}\label{lem:solgenDSE}
The Dyson-Schwinger equation (\ref{eq:genDSE}) has a unique solution
described by $c_0=\One$ and
\begin{equation}\label{eq:genreccn}
c_{n} = \sum_{m=1}^{n} w_{m}B_+^{d_{m}}\left(\sum
_{k_1+\ldots+k_{m+1}=n-m,\, k_i\ge 0} c_{k_1}\ldots
c_{k_{m+1}}\right).
\end{equation}
\end{lem}
\emph{Proof. } Inserting the ansatz into (\ref{eq:genDSE}) and
sorting by powers of $\alpha$ yields the result. Uniqueness is
obvious. $\quad\Box$
\begin{thm}\label{thm:dsesub}
 The elements $c_n$ generate a Hopf subalgebra of $A:$
\begin{equation*}
\Delta(c_n)= \sum_{k=0}^n P^n_k\otimes c_k
\end{equation*}
where the $P^n_k$ are homogeneous polynomials of degree $n-k$ in the
$c_l,$ $l\le n:$
\begin{equation}\label{eq:pnk}
P^{n}_k = \sum_{l_1+\ldots+l_{k+1}=n-k}c_{l_1}\ldots c_{l_{k+1}}.
\end{equation}
In particular, the $P^n_k$ are independent of the $w_n$ and
$B_+^{d_n}.$
\end{thm}
We emphasize that the main ingredient for the proof of this theorem
is the fact that the $B_+^{d_n}$ are Hochschild 1-cocycles, the rest
being a cumbersome but
straightforward calculation. \\\\
\emph{Proof. }We proceed by proving inductively the following
statements:
\begin{enumerate}
\item[$(\alpha_n)$] The theorem holds up to order $n.$
\item[$(\beta_n)$] For a given $m\in\{1\ldots n\}$ let $l_1+\ldots+l_{m+1}=:p\in\{0\ldots n-m\},$ $l_i\ge 0.$ Then the right hand sum
\begin{equation}\label{eq:notbn}
P(n-m,m,p):=\sum_{k_1+\ldots+ k_{m+1}=n-m,\,k_i\ge l_i}
P^{k_1}_{l_1}\ldots P^{k_{m+1}}_{l_{m+1}}
\end{equation}
does not depend on the single $l_i$ but only on $p,$ $n-m$ and $m,$
justifying the notation $P(n-m,m,p).$
\item[$(\gamma_n)$] In the above notation and for any $q\in\{1\ldots n\}$, the term $P(n-m,m,q-m)$ does not depend
on $m\in\{1\ldots q\}.$
\end{enumerate}
To start the induction, we note that $(\alpha_0)$ is obvious.
$(\beta_1)$ is trivial as $m=1$ enforces $l_1=l_2=0.$ Similarly,
for $(\gamma_1)$ only one $m$ is in range and the statement thus
trivially satisfied. We proceed to $(\alpha_n).$ By definition,
and using (\ref{eq:Hscb1}) for the $B_+^{d_n},$
\begin{eqnarray*}
\Delta(c_n) &=& \sum_{m=1}^n w_m ((id \otimes
B_+^{d_m})\Delta+B_+^{d_m}\otimes\One)\\
&&
\left(\sum _{k_1+\ldots+k_{m+1}=n-m,\,
k_i\ge 0} c_{k_1}\ldots c_{k_{m+1}}\right)
\end{eqnarray*}
(using the induction hypothesis $(\alpha_{n-1})$)
\begin{eqnarray*}
&=&c_n\otimes\One+\sum_{m=1}^n w_m (id \otimes B_+^{d_m})\sum
_{k_1+\ldots+k_{m+1}=n-m,\, k_i\ge 0}\sum_{l_1\ldots
l_{m+1}=0}^{k_1\ldots
k_{m+1}}\\
&&P^{k_1}_{l_1}\ldots P^{k_{m+1}}_{l_{m+1}}\otimes c_{l_1}\ldots
c_{l_{m+1}}=
\end{eqnarray*}
(by rearranging indices)
\begin{eqnarray*}
&=&c_n\otimes\One+\sum_{m=1}^n w_m \sum_{p=0}^{n-m}\sum_{l_1+\ldots
+l_{m+1}=p}\sum
_{k_1+\ldots+k_{m+1}=n-m,\, k_i\ge l_i}\\
&&P^{k_1}_{l_1}\ldots P^{k_{m+1}}_{l_{m+1}}\otimes
B_+^{d_m}(c_{l_1}\ldots c_{l_{m+1}})=
\end{eqnarray*}
(by the induction hypothesis $(\beta_{n})$ and using the notation
of (\ref{eq:notbn}))
\begin{eqnarray*}
&=&c_n\otimes\One+\sum_{m=1}^n w_m
\sum_{p=0}^{n-m}P(n-m,m,p)\otimes \sum_{l_1+\ldots +l_{m+1}=p}
B_+^{d_m}(c_{l_1}\ldots c_{l_{m+1}})=
\end{eqnarray*}
(rearranging indices ($q$ replaces $m+p$) and using
$(\gamma_{n})$)
\begin{eqnarray*}
&=&c_n\otimes\One+\sum_{q=1}^n
\sum_{m=1}^{q}w_mP(n-m,m,q-m)\otimes\\
&&\otimes
\sum_{l_1+\ldots +l_{m+1}=q-m} B_+^{d_m}(c_{l_1}\ldots c_{l_{m+1}})=\\
&=&c_n\otimes\One+\sum_{q=1}^n P(n-q,q,0)\otimes \sum_{m=1}^q
w_q\sum_{l_1+\ldots +l_{m+1}=q-m} B_+^{d_m}(c_{l_1}\ldots
c_{l_{m+1}}).
\end{eqnarray*}
Since the right hand tensor factor is $c_q,$ a glance at (\ref{eq:notbn}), using that $P^k_0=c_k,$ verifies $(\alpha_n).$\\\\
The items $(\beta_n)$ and $(\gamma_n)$ follow from
$(\alpha_{n-1}):$
\begin{eqnarray*}
P(n-m,m,p)&=&\sum_{k_1+\ldots+ k_{m+1}=n-m,\,k_i\ge l_i}
P^{k_1}_{l_1}\ldots P^{k_{m+1}}_{l_{m+1}}=\\
&=&\sum_{k_1+\ldots+ k_{m+1}=n-m,\,k_i\ge l_i}
\sum_{r_1^1+\ldots+r^1_{l_1+1}=k_1-l_1}\ldots\\
&&\ldots
\sum_{r_1^{m+1}+\ldots+r^{m+1}_{l_{m+1}+1}=k_{m+1}-l_{m+1}}
c_{r_1^1}\ldots c_{r^{m+1}_{l_{m+1}+1}}=\\
&=&\sum_{r_1+\ldots+ r_{m+p+1}=n-m-p}c_{r_1}\ldots c_{r_{m+p+1}},
\end{eqnarray*}
which is independent of any $l_i$ whence $(\beta_n).$ Substituting $p=q-m$ shows $(\gamma_n).$ \hfill $\Box$ \\\\
At first sight the fact that the coproduct on the $c_i$ does not
depend on the $w_k$ and hence that all Dyson-Schwinger equations of
this kind yield isomorphic Hopf subalgebras (provided there are no
relations among the $c_n$) might well come as a surprise. The deeper
reason for this is the recursiveness of (\ref{eq:genDSE}) as will
become more apparent in the next paragraphs.
\subsubsection*{Description in terms of trees}
Now we specialize to the case $A=\mathcal{H}(S)$ where $(S=\dot\cup
S_n,|\cdot|)$ is an arbitrary graded set of decorations such that
$|d_n|=n$ for all $n$ (one can even allow $d_n\subset S_n$ and
define $B_+^{d_n}:= \sum_{\delta\in d_n} B_+^\delta).$ The maps
$B_+^{d_n}$ are defined as in (\ref{eq:defbplus}) where the newly
created vertex is decorated by $d_n.$ Using the following lemma,
which gives an explicit presentation of the $c_i$ in terms of trees,
\tref{thm:dsesub} can be proven in a more comprehensive way.
\begin{lem}\label{lem:solDSEtrees} The solution of (\ref{eq:genDSE}) can be described by $c_0=\One$ and
\begin{equation}\label{eq:gencn}
c_{n}=\sum_{\tau\in\mathcal{T}(S),\, |\tau|=n}
\frac{\tau}{\operatorname{Sym}(\tau)}\prod_{v\in\tau^{[0]}} \gamma_v
\end{equation}
where
\begin{equation*}
\gamma_v= \left\{ \begin{array}{rl}
w_{|dec(v)|}\frac{(|dec(v)|+1)!}{(|dec(v)|+1-\operatorname{fert}(v))!}
&\mbox{ if }\operatorname{fert}(v)\le |dec(v)|+1 \\ 0 & \mbox{
else}.
\end{array} \right.
\end{equation*}
Here $\mathcal{T}(S)$ denotes the set of $S$-decorated trees,
$\tau^{[0]}$ the set of vertices of $\tau,$ $dec(v)$ the decoration
(in $S$) of $v,$ $|\tau|$ the decoration weight of $\tau,$ i.~e.~
$|\tau|=\sum_{v\in\tau^{[0]}} |dec(v)|,$ and
$\operatorname{fert}(v)$ the fertility (number of outgoing edges) of
the vertex $v.$
\end{lem}
\emph{Proof. }This is an easy induction using the following
argument: Let $\tau$ be a given tree in $c_n$ and let its root $o$
be decorated by something in degree $m.$ According to
(\ref{eq:genreccn}), $\tau=B_+^{d_m}(\One^{k_0}\tau_1\ldots
\tau_{m+1-k_0})$ where the $\tau_i$ are trees different from
$\One.$ The fertility of the root is thus $m+1-k_0.$ We assume
$\tau_1\ldots\tau_{m+1-k_0}=\sigma_1^{k_1}\ldots\sigma_p^{k_p}$
where the $\sigma_i$ are pairwise different trees. In
(\ref{eq:genreccn}), there are $C := \frac{(m+1)!}{k_0!\ldots
k_p!}$ choices to make which yield the tree $\tau.$ Since the
$\gamma_v$ are simply multiplied for all vertices $v$ of a tree,
it remains to see that for the only new vertex $o$ in $\tau,$ we
have
\begin{equation*}
\gamma_o/w_m=\frac{(m+1)!}{k_0!}=C\frac{\mbox{Sym}(\tau)}{\mbox{Sym}(\tau_1)\ldots\mbox{Sym}(\tau_{m+1-k_0})}.
\end{equation*}
This however follows immediately from the definition of
Sym.\hfill$\Box$
\\\\
As a matter of fact, the coefficients
\begin{equation}\label{eq:dsecoeff}
\prod_v \gamma_v
\end{equation}
can be interpreted as follows: Consider each tree as an
''operadic`` object with $|dec(v)|+1-\operatorname{fert}(v)$
inputs at each vertex $v.$ For example,
\begin{equation*}
\oforknum \quad \rightarrow\quad \oforknumhair
\end{equation*}
Clearly, the total number of inputs is $n+1$ for any tree of
decoration weight $n.$ Now the coefficient (\ref{eq:dsecoeff}) is
nothing but the number of planar embeddings of this operadic tree
(where the trunk, i.~e.~ the original tree is kept fixed). In other
words, (\ref{eq:dsecoeff}) counts the number of ways that the input
edges can sway around the original tree. Using this idea, we obtain
the following
\subsubsection*{Operadic proof of \tref{thm:dsesub}}
As a variation of (\ref{eq:genDSE}) let us consider the operadic
fixpoint equation
\be G(\a)= \One + \sum_n \a^n \mu_{n+1}(G(\a)^{\otimes (n+1)}). \ee
Here, $\mu_j$ is a map $\in O^{[j]}: V^{\otimes j}\to V$ for some
space $V$ and $G(\a)$ is a formal series in $\a$  with
coefficients in the $O^{[j]}.$ We regard $\One: V\to V$ as the
identity map. We write $G(\a)=\One+\sum_k \a^k \nu_{k}$. It
follows easily by induction that $\nu_{k}\in \;O^{[k+1]}.$
Clearly, $G(\a)$ is a sum (with unit weights) over all maps which
we obtain
by composition of some undecomposable maps $\mu_n$.\\\\
The coproduct of decorated rooted trees acts on the $\nu_k$ in an
obvious manner. A given monomial $\nu_{i_1}^{r_1}\cdots
\nu_{i_l}^{r_l}$ (which lives in the PROP $V^{\otimes
(r_1i_1+\ldots+r_li_l+r)}\to V^{\otimes r}$, where $r=\sum r_i$)
can be composed with any element in $O^{[r-1]}$ in
\begin{equation}\label{eq:factorial}
\frac{r!}{r_1!\ldots r_l!}
\end{equation}
ways. Hence, as the $\nu_i$ sum over all maps with unit weight,
this is the contribution to the term in the coproduct which has
$\nu_{k}$ on the right hand side and the given monomial on the
left hand side. Going back to the initial Dyson-Schwinger equation
(\ref{eq:genDSE}), we see that the same argument
(\ref{eq:factorial}) also determines the coproduct on the $c_k$
there, in agreement with (\ref{eq:pnk}):
\begin{equation}\label{eq:dsecoeffagain}
P^n_k = \sum_{\genfrac{}{}{0pt}{}{i_1r_1+\ldots+i_lr_l=n-k}{0\le
i_s<i_{s+1},\,\sum r_i=k+1}}\frac{(k+1)!}{r_1!\ldots
r_l!}c_{i_1}^{r_1}\ldots c^{r_l}_{i_{l}}
\end{equation}
Indeed, the trees in $c_k$ were weighted by a product over vertices
(\ref{eq:dsecoeff}), and the coproduct respects the
planar structure. \hfill $\Box$\\\\
The coefficients in (\ref{eq:dsecoeff}) and (\ref{eq:dsecoeffagain})
arise thus in a completely natural way due to the transition from a
noncommutative (planar) to a commutative (non-planar) setting.
\subsubsection*{Final remarks}
Before we ultimately turn to the more analytical side of
Dyson-Schwinger equations, let us mention that by \tref{thm:dsesub},
the Connes-Moscovici Hopf subalgebra presented in the preceding
subsection is not generated by a Dyson-Schwinger equation of the
form (\ref{eq:genDSE}) if we restrict ourselves to one-cocycles
mapping into the linear space of generators, as they
typically appear in local quantum field theory. \\\\
Note that the Hopf algebras which appear as solutions of
(\ref{eq:genDSE}) are studied under the name Faa di Bruno algebras
in \cite{FGB}, to which the Connes-Moscovici algebra can be
related through an isomorphism.\\\\
In studying propagator insertions, one encounters the more general
equation \begin{equation*} X=\alpha
B_+\left(\frac{\One}{\One-X}\right).
\end{equation*}
It yields a Hopf subalgebra in a similar way, see \cite{brkdse} for details. \\\\
Finally let us emphasize that the ladder Hopf algebra
$\mathcal{H}_1$ introduced in the last subsection, can be
generated by the \emph{linear} Dyson-Schwinger equation
\begin{equation*}
X=\One+\alpha B_+(X)
\end{equation*}
The Hopf algebra $\mathcal{H}_1$ plays a special role at the
fixpoint of the renormalization group flow \cite{MK2}, see also
the
next subsection.\\\\
As opposed to the above example, we call Dyson-Schwinger equations
of the form (\ref{eq:genDSE}) (where some $w_n\neq 0$ for $n\ge 1$)
\emph{nonlinear}. They necessarily generate trees with
sidebranchings.
\subsection{Applications in physics and number theory}
In physics, Dyson-Schwinger equations, usually derived by formal
means using functional integrals, describe the loop expansion of
Green functions in a recursive way. An alternative to derive these
equations is given by the very existence of a Hopf algebra
underlying perturbation theory. These Hopf algebras provide
Hochschild 1-cocycles, and we can obtain the Dyson--Schwinger
equations for them in a straightforward manner.\\\\
In the following, we first exhibit Dyson-Schwinger equations in
three different contexts: as a source for transcendental numbers, as
a manner to define a generating function for the polylogarithm, and
as the equations of motion for a renormalizable quantum field
theory. The presentation is by no means self-contained, and we refer
the reader to the growing literature for more details
\cite{annals,dennisfest,banz,numberswe,zinnowitz,inprep}.
\subsubsection*{A simple toy model}
Let us consider the equation we had before (\ref{eq:quadrdse}), \bel
X_2=\One+\alpha B_+(X_2^2),\label{eq:dseq}\eel and let us exhibit
the difference between such an equation and the associated linear
system \bel X_1=\One+\alpha B_+(X_1).\label{eq:dsel}\eel We will
study toy Feynman rules on these Hopf algebras, regarded as
characters on the Hopf algebra. We explore that $\phi B_+(\One)$
defines an integral kernel $k$, such that \be \phi
B_+(\One)[z]=\int_0^\infty k(x,z)dx,\ee where the kernel is
homogeneous: $k(ux,uz)=k(x,z)/u$. We regard the integral as the
Fourier transform of the kernel with respect to the multiplicative
group $\mathbb{R}_+$. To define our first set of renormalized
Feynman rules, we simply set \be \phi B_+(h)[z]=\int_0^\infty
(k(x,z)-k(x,1))\phi(h)[x] dx.\ee Note that we have
$\phi(h_1h_2)=\phi(h_1)\phi(h_2)$ which implies
$\phi(\One)[z]=1$.\\\\
Let us define the transform $K(\gamma)$ of the kernel $k(x,1)$ to
be \be K(\gamma)=\int_0^\infty k(x,1)x^{-\gamma}dx.\ee This
determines \be \int_0^\infty
k(x,z)x^{-\gamma}dx=z^{-\gamma}K(\gamma).\ee \\\\
Let us now look at (\ref{eq:dsel}). Applying $\phi$ to both sides
delivers an integral equation for the Green function \be
\phi(X_1)[z;\alpha]=1+\alpha \int_0^\infty
\phi(X_1)[x;\alpha](k(x,z)-k(x,1))dx.\ee Note that our choice of
renormalized Feynman rules corresponds to the choice of a boundary
condition for the Dyson-Schwinger equation,
$\phi(X_1)[1;\alpha]=1$. Omitting the subtraction of the kernel at
$z=1$ defines the unrenormalized Feynman rule $\phi_u$, which
reconstructs the renormalized one, $\phi=S_R^{\phi_u}\star \phi_u
$, where $R$ is the evaluation map at $z=1$. \\\\
Equation (\ref{eq:dsel}) can be solved by an Ansatz
$\phi(X_1)[z]=z^{-\gamma(\alpha)}$, which leads to \be
z^{-\gamma(\alpha)}=1+\alpha
(z^{-\gamma(\alpha)}-1)K(\gamma(\alpha)),\ee i.~e.\ the series
$\gamma(\alpha)$ is the solution of the equation $1=\alpha
K(\gamma(\alpha))$. The non-linear case (\ref{eq:dseq}) cannot be
solved by such an {\it Ansatz}. Maintaining the same boundary
condition we get the equation \be \phi(X_2)[z;\alpha]=1+\alpha
\int_0^\infty (\phi(X_2)[x;\alpha])^2(k(x,z)-k(x,1))dx.\ee \\\\
At this moment, it is instructive to introduce a bit of quantum
field theory wisdom. We observe that we can write this integral
equation in the form \be \phi(X_2)[z;\alpha]=1+ \int_0^\infty
\phi(X_2)[x;\alpha]\beth(x;\alpha)(k(x,z)-k(x,1))dx,\ee where the
running coupling $\beth(x;\alpha)=\alpha \phi(X_2)[x;\alpha]$ has
been introduced. We see that we just modify the linear
Dyson-Schwinger equation by this running coupling, which forces us
to look for solutions not of the form \be
G(\alpha,z)=e^{-\gamma(\alpha)},\ee but instead of the more
general form \bel
G(\alpha,z)=e^{-\sum_{j=1}^\infty\gamma_j(\alpha)\ln^j
z},\label{eq:recurs}\eel where the $\gamma_j$ themselves are
recursively defined through $\gamma_1$ thanks to the
renormalization group. \\\\
Indeed, assume now that the running coupling is  constant,
$\partial_{\ln z}\beth=0$. This turns the non-linear
Dyson-Schwinger equation into the linear one (\ref{eq:dsel}). That
is a general phenomenon: the linear Dyson-Schwinger equation
appears in the limit of a vanishing $\beta$-function, and
signifies a possible fixpoint of the renormalization group.
\\\\
This suggests a natural expansion in terms of the coefficients of
the $\beta$-function, which will be presented elsewhere.\\\\
All this has a combinatorial counterpart: \be \frac{\partial \ln
X_2(\alpha)}{\partial\alpha}=S\star Y(X_2(\alpha_R)),\ee where $Y$
is again the grading operator. Let us work this out in an example.
We consider the solution $X_2(\alpha)$ of (\ref{eq:dseq}). Setting
$X_2=\One+\sum_{k=1}^\infty \alpha^k c_k$, we find to $O(\alpha^3)$,
\be
\begin{array}{ll}\tilde\Delta(c_1)=0 & S*Y(c_1)=c_1\\
\tilde\Delta(c_2)=2c_1\otimes c_1 & S*Y(c_2)=2c_2-2c_1^2\\
\tilde\Delta(c_3)=3c_1\otimes c_2+(2c_2+c_1^2)\otimes c_1 &
S*Y(c_3)=3c_3-8c_1c_2+5c_1^3.\end{array} \ee
%
%
Furthermore, \be \alpha\partial_\alpha \ln X_2(\alpha)=\alpha
c_1+2\alpha^2(c_2-\frac{1}{2}c_1^2)+3\alpha^3\left(c_3-c_1c_2+\frac{1}{3}c_1^3\right).\ee
Setting \be \alpha=\frac{\alpha_R}{X(\alpha)},\ee and recursively
replacing $\alpha$ by $\alpha_R$, \beas \alpha(\alpha_R) & = &
\frac{\alpha_R}{1+\alpha(\alpha_R)c_1+\alpha^2(\alpha_R)c_2+\cdots}\\
& = & \frac{\alpha_R}{1+\frac{\alpha_R
c_1}{1+\alpha_Rc_1+\cdots}+\alpha_R^2c_2+\cdots}\\
& = & \frac{\alpha_R}{1+\alpha_R
c_1+\alpha_R^2(c_2-c_1^2)+\cdots},\eeas confirms the result to
that order. The general proof is a straightforward application of
the results in \cite{RHII}. \\\\
Furthermore, the reader can check that \be F_{c_m}(c_k)=(k-m+1)
c_{k-m}\ee for all $k\geq m$, which is at the heart of a recursive
determination of the above coefficients $\gamma_j(\alpha)$ in
(\ref{eq:recurs}). Here, $F_{c_m}$ is the befooting operator \be
F_{c_m}(c_k)=\left<Z_{c_m}\otimes {\rm id},\Delta(c_k)\right >\ee of
\cite{brkdse}.
As a final remark, we mention that it is not the non-linearity
which provides the major challenge in solving a non-linear
Dyson-Schwinger equation, but the fact that the one-variable
Fourier calculus presented above has to be replaced by a
multi-variable calculus which leads to transcendental extensions
\cite{zinnowitz} which is a fascinating
topic in its own right.\\\\
Indeed, consider once more the linear equation (\ref{eq:dsel}),
now with Feynman rules defined by a two-variable kernel
$k(x,y,z)=1/(x+y+z)^2 $ with $k(xz,yz,z)=k(x,y,1)/z^2 $ \be
\phi(B_+(h))(z)=\int dxdy
\frac{(\phi(h)(x))^{q_2}(\phi(h)(y))^{q_1}}{(x+y+z)^2}-|_{z=1},\ee
and $q_1,q_2$ are two positive rational numbers which add to one.
Comparing this system with the degenerate system where one of the
$q_i$ vanishes (and the other thus is unity) shows that the
perturbative expansion in $\alpha$ provides coefficients which are
transcendental extensions of the ones obtained in the degenerate
case. This rather general phenomenon leads deeply into the
transcendental structure of Green functions, currently under
investigation.
\subsubsection*{Dyson-Schwinger equation for the polylog}
Quantum field theory is concerned with the determination of
correlators which we can regard as generating functions for a
perturbative expansion of amplitudes. These correlators are
solutions of our Dyson-Schwinger equations, the latter being typical
fixpoint equations: the correlator equals a functional of the
correlator. Such self-similarities appear in many branches of
mathematics. Here, we want to exhibit one such appearance which we
find particularly fascinating: the generating function for the
polylog \cite{numberswe}. \\\\
Following \cite{numberswe} consider the following $N\times N$
matrix once more borrowed from Spencer Bloch's function theory of
the polylogarithm \cite{Bloch}: \be
\begin{array}{c}
  \alpha^0 \\
  \alpha^1 \\
  \alpha^2 \\
  \alpha^3 \\
  \ldots
\end{array}
\underbrace{\left(
\begin{array}{rcrcrcrcr}
  + 1 & \mid & 0 & \mid & 0 & \mid & 0 & \mid & \cdots\\
  -{\rm Li}_1(z) & \mid & 2\pi i & \mid & 0 & \mid & 0 & \mid & \cdots \\
  -{\rm Li}_2(z)  & \mid & 2\pi i \ln z & \mid & [2\pi i]^2 & \mid & 0 & \mid &  \cdots\\
  -{\rm Li}_3(z)  & \mid & 2\pi i \frac{\ln^2 z}{2!} & \mid & [2\pi i]^2\ln z & \mid & [2\pi i]^3 & \mid &  \cdots\\
  \ldots & \mid & \ldots & \mid & \ldots & \mid & \ldots & \mid & \ldots
\end{array}\right)}_{\hfill u^0\hfill u^1\hfill u^2\hfill u^3\hfill\ldots},
\ee given up to $N=4$. We assign an order in a small parameter
$\alpha$ to each row, counting rows $0,1,\ldots$ from top to
bottom, similarly we count columns $0,1,\ldots$ from left to right
by a parameter $u$, and assign an order $u^i$ to the $i$-th
column. The polylog is defined by \be {\rm
Li}_n(z)=\sum_{k=1}^\infty\frac{z^k}{k^n} \ee inside the unit
circle and analytically continued with a branch cut along the real
axis from one to plus infinity. \\\\
As the $n$-th polylog appears as the integral over the $(n-1)$-th
polylog, we expect to be able to find a straightforward integral
equation for its generating function which resembles a
Dyson-Schwinger equation. Consider \beas F(\alpha,u;z) & = &
1-\frac{1}{1-z}+\frac{2\pi i u \alpha}{1-2\pi i u
\alpha}+\alpha\left[ \int_0^z
\frac{F(\alpha,0;x)}{x}dx\right.\nonumber\\ & & \left. +\int_1^z
\frac{F(\alpha,u;x)-F(\alpha,0;x)}{x}dx\right],\eeas where we call
$F(\alpha,u;z)$ a renormalized Green function, $\alpha$ the coupling
(a small parameter, $0<\alpha<1$) and consider the perturbative
expansion \be F(\alpha,u;z) =1-\frac{1}{1-z}+\sum_{k=1}^\infty
\alpha^k f_k(u;z).\ee We distinguished the lowest order term
$f_0(z)=z/(z-1)$ (which corresponds to the term without quantum
corrections in QFT) at order $\alpha^0$ which here equals $-{\rm
Li}_0(z)$. The limit $u\to 1$ can be taken in the above
Dyson-Schwinger equation. We note that upon introducing a
counterterm $Z(\alpha,u;\ln \rho)$, the above equation is the
renormalized solution at $\rho\to 0$ of the equation \be
F_\rho(\alpha,u;z)=Z(\alpha,u;\ln \rho)-\frac{1}{1-z}+\frac{2\pi i u
\alpha}{1-2\pi i u \alpha}+\alpha \int_0^z
\frac{F_\rho(\alpha,u;x)}{x}dx.\ee
\\\\
We immediately confirm that, for $k>0$, the term of order $\alpha^k
u^i$ in the renormalized solution of this Dyson-Schwinger equation
is the entry $(k,i)$ in the above matrix: the above matrix provides
in its non-trivial entries the solution of the
Dyson-Schwinger equation so constructed. \\\\
We now work with the cocommutative Hopf algebra $\mathcal{H}_1$
determined by the Dyson-Schwinger equation $X=\One+\alpha B_+(X)$,
so $X=\sum_{k=0}^\infty \alpha^k t_k$ where the $t_k$ are $k$-fold
application of $B_+$ to $\One$, and let $Li\equiv Li(z)$ and
$L\equiv L(z)$ be characters on the Hopf algebra defined by \be
-\phi(t_n)(z,0)\equiv Li(t_n)(z)={\rm
Li}_n(z),\;L(t_n)(z)=\frac{\ln^n(z)}{n!}.\ee We can regard the
character $Li$ as a Feynman rule and the transition $Li\to L$ as a
renormalization map which leaves the behavior at infinity
unchanged. \\\
We know \cite{Bloch} that the elimination of all ambiguities due to
a choice of branch lies in the construction of functions
$a_p(z)=(2\pi i)^{-p}\;\widetilde{a}_p(z)$ where \be
\widetilde{a}_p(z):= {\rm Li}_p(z)-\cdots+ (-1)^j {\rm
Li}_{p-j}(z)\frac{\ln^j(z)}{j!}+\cdots +(-1)^{p-1}{\rm
Li}_1(z)\frac{\ln^{p-1}(z)}{(p-1)!}.\ee This is now a very familiar
equation:
\begin{pro} For $z\in\C$, \be \tilde{a}_p(z)=m
(L^{-1}\otimes Li)({\rm id}\otimes P)\Delta(t_p),\ee where
$L^{-1}=LS$, with $S$ the antipode in $\mathcal{H}_1,$ and $P$ the
projection onto the augmentation ideal.
\end{pro}
\emph{Proof:} elementary combinatorics confirming that \be LS(t_n/n!)(z)=(-\ln(z))^n/n!. \ee \hfill $\Box$ \\\\
There is a strong analogy here to the Bogoliubov $R$ operation in
renormalization theory \cite{annals,numberswe}, thanks to the fact
that $Li$ and $L$ have matching asymptotic behavior for $\mid\!
z\!\mid\to\infty$. Indeed, if we let $R$ be defined to map the
character $Li$ to the character $L$, $R(Li)=L$, and $P$ the
projector onto the augmentation ideal of $\mathcal{H}_1$, then \be
LS=S_R^{Li}=-Rm (S^{Li}_R\otimes Li)({\rm id}\otimes P)\Delta\equiv
-R\left(\overline{Li}\right), \ee for example \be
S_R^{Li}(t_2)=-R(Li(t_2)+S_R^{Li}(t_1)Li(t_1))=-L(t_2)+L(t_1)L(t_1)=\frac{\ln^2(z)}{2!},\ee
where $\overline{Li}(t_2)=Li(t_2)-L(t_1)Li(t_1)$. Thus, $a_p$ is the
result of the Bogoliubov map \be\overline{Li}=m(S^{Li}_R\otimes
Li)({\rm id}\otimes P)\Delta\label{prop1}\ee acting on $t_n$.
We have two completely equivalent mechanism for the removal of
ambiguities at this moment: \be L^{-1}\star Li\,\mbox{ vs }\,
S_R^\phi\star\phi.\ee This points towards an analogy between the
structure of the polylog and QFT Green functions which very much
suggests to explore QFT from the viewpoint of mixed Hodge
structures in the future.
\subsubsection*{Dyson-Schwinger equations for full QFT}
The quantum equations of motion, the Dyson-Schwinger equations of
a full fledged quantum field theory, can be obtained in precisely
the same manner as discussed above.  They typically are of the
form \bel \Gamma^{\underline{r}}= \One +
\sum_{\genfrac{}{}{0pt}{}{\gamma\in H_L^{[1]}}{{\rm
res}(\gamma)=\underline{r}}} \frac{\alpha^{\vert\gamma\vert}}{{\rm
Sym}(\gamma)} B_+^\gamma(X_{\mathcal
R}^\gamma)=\One+\sum_{\genfrac{}{}{0pt}{}{\Gamma\in H_L}{ {\rm
res}(\Gamma)=\underline{r}}}\frac{\alpha^{\vert\Gamma\vert}\Gamma}{{\rm
Sym}(\Gamma)}\; , \label{eq:dseqft}\eel
 where the first sum is over a
countable set of Hopf algebra primitives $\gamma$, ${\rm
res}(\gamma)=\underline{r}$, \be \Delta(\gamma)=\gamma\otimes\One
+ \One\otimes \gamma,\ee indexing the Hochschild 1-cocycles
$B_+^{\gamma}$ above, while the second sum is over all
one-particle irreducible graphs contributing to the desired Green
function, all weighted by their symmetry factors. In more
traditional terms, the primitive graphs $\gamma$ correspond to
skeletons into which vertex and propagator corrections are to be
inserted.\\\\
Here, $\Gamma^{\underline{r}}$ is to be regarded as
 a formal series
\be  \Gamma^{\underline{r}}=\One+\sum_{k\geq 1}
c_k^{\underline{r}} \alpha^k, \;c_k^{\underline{r}}\in H.\ee
These coefficients of the perturbative expansion deliver Hopf
subalgebras in their own right, cf.~\tref{thm:dsesub}. Indeed, the
maps \be B_+^{\underline{r},n}=\sum_{\genfrac{}{}{0pt}{}{\gamma\in
H_L^{[1]}}{{\rm
res}(\gamma)=\underline{r},\;|\gamma|=n}}B_+^\gamma,\ee where the
sum is over all primitive 1PI $n$-loop graphs $\gamma$ with
external leg structure $\underline{r}$, are 1-cocycles. They are
implicitly defined by the second equality in (\ref{eq:dseqft}),
the remarkable feature is the fact that these maps can be shown to
be Hochschild closed and hence ensure locality. A detailed account
of this fact, which illuminates in particular the
structure of gauge theories, is upcoming \cite{inprep}. \\\\
In (\ref{eq:dseqft}), $X_{\mathcal R}^\gamma$ is of the form \be
X_{\mathcal R}^\gamma=\Gamma^{{\rm res}(\gamma)}\left(X_{\rm
coupl}\right)^{|\gamma|},\ee where $X_{\rm coupl}$ is the vertex
function divided by the square roots of the inverse propagator
functions. Under the Feynman rules $X_{\rm coupl}$ hence maps to
the
invariant charge.\\\\
%
%
As an example, consider QED. We have a set of residues (external
leg structures) \be {\rm Res}=\left\{\qedzggb,\qedzff,\qedzgff
\right\}.\ee We finish our paper by exhibiting the action of the
Hochschild 1-cocycle $B_+^{\qedagffa}$ on the order $\alpha$
expansion of \be
X_{\qedagffa}=\frac{\left(\Gamma^{\qedzgff}\right)^3}{\left(\Gamma^{\qedzff}\right)^2\left(\Gamma^{\qedzggb}\right)}=\Gamma^{\qedzgff}
\left(X_{\rm coupl}\right)^2,\ee with \be X_{\rm
coupl}=\Gamma^{\qedzgff}\left(\Gamma^{\qedzff}\sqrt{\Gamma^{\qedzggb}}\right)^{-1}.\ee
\\\\
To order $\alpha$, one finds \be X_{\qedagffa}=1+\alpha\left(
3\qedagffa+2\qedaffa+\qedagga.\right)\ee Hence, the non-primitive
two-loop vertex graphs of QED are obtained as \be
B_+^{\qedagffa}\left( 3\qedagffa+2\qedaffa+\qedagga\right).\ee
Hochschild closedness demands that this equals \be
\qedbfgga+\qedbfggb+\qedbfggc+\qedbfggd+\qedbfgge+\qedbfggf,\ee as
then \be \tilde\Delta(\ldots)=
\left(3\qedagffa+2\qedaffa+\qedagga\right)\otimes \qedagffa.\ee In
this manner one determines the Hochschild 1-cocycles for a
renormalizable quantum field theory. This works particularly nice
for gauge theories, as will be exhibited in \cite{inprep}.
%
\subsection{Final remarks}
There is a very powerful structure behind  the above decomposition
into Hopf algebra primitives -- the fact that the sum over all
Green functions $G^{\underline{n}}$ is indeed the sum over all 1PI
graphs, and this sum, the effective action, can be written nicely
as $\prod\frac{1}{1-\gamma}$, a product over "prime" graphs --
graphs which are primitive elements of the Hopf algebra and which
index the Hochschild 1-cocycles, delivering a complete
factorization of the action. A single such Euler factor with its
corresponding Dyson-Schwinger equation and Feynman rules was
evaluated in \cite{brkdse}, a calculation which was entirely in
accordance with our study: an understanding of the weight of
contributions $\sim \ln(z)$ from a knowledge of the weight of such
contributions of smaller degree in $\alpha$, dubbed
propagator-coupling duality in \cite{brkdse}. Altogether, this
allows to summarize the structure in QFT as a vast generalization
of results summarized here. It turns out that even the quantum
structure of gauge theories can be understood along these lines
\cite{Kreimer2}. A full discussion is upcoming \cite{inprep}.
%
%
%

\bibliographystyle{plain}
\bibliography{lit}

\end{document}